\documentclass[12pt]{article}

\usepackage{a4wide}
\usepackage{amsmath,amssymb}
\usepackage{slashed}
\usepackage{xcolor}
\usepackage{graphicx}
\usepackage{url}
\usepackage{cite}
\usepackage[colorlinks=true,allcolors=darkred,pdfborder={0 0 0}]{hyperref}

\def\rpv{$\slashed{R}_p \:$}
\def\rpvn{$\slashed{R}_p$}
\definecolor{darkred}{rgb}{0.6,0,0}

\begin{document}

\vspace*{-2cm}
\begin{flushright}
IFIC/15-16 \\
\vspace*{2mm}
\today
\end{flushright}

\begin{center}
\vspace*{15mm}

\vspace{1cm}
{\Large \bf 
Lepton flavor violation beyond the MSSM
} \\
\vspace{1cm}

{\bf A. Vicente$^{a,b}$}

 \vspace*{.5cm} 
$^{a}$ Instituto de F\'{\i}sica Corpuscular (CSIC-Universitat de Val\`{e}ncia), \\
Apdo. 22085, E-46071 Valencia, Spain
\vspace*{.2cm} 

$^{b}$ IFPA, Dep. AGO, Universit\'e de Li\`ege, \\
Bat B5, Sart-Tilman B-4000 Li\`ege 1, Belgium

\end{center}

\vspace*{10mm}
\begin{abstract}\noindent\normalsize
Most extensions of the Standard Model lepton sector predict large
lepton flavor violating rates. Given the promising experimental
perspectives for lepton flavor violation in the next few years, this
generic expectation might offer a powerful indirect probe to look for
new physics. In this review we will cover several aspects of lepton
flavor violation in supersymmetric models beyond the Minimal
Supersymmetric Standard Model. In particular, we will concentrate on
three different scenarios: high-scale and low-scale seesaw models as
well as models with R-parity violation. We will see that in some cases
the LFV phenomenology can have characteristic features for specific
scenarios, implying that dedicated studies must be performed in order
to correctly understand the phenomenology in non-minimal
supersymmetric models.
\end{abstract}

\newpage

\tableofcontents

\newpage

\section{Introduction} \label{sec:intro}

The Standard Model (SM) particle content has been recently completed
with the discovery of the long-awaited Higgs boson at the CERN Large
Hadron Collider (LHC)~\cite{Aad:2012tfa,Chatrchyan:2012ufa}. This
constitutes a well deserved reward after decades of intense search,
with great efforts from the theory and experimental
communities. Furthermore, it also confirms that the SM must be, at
least to a good approximation, a precise description of nature up to
the energies explored. In fact, and apart from some phenomenological
facts that indeed require some unknown new physics (NP), like the
existence of dark matter and neutrino masses, the SM explains to a
high level of accuracy all the observations made in a wide variety of
experiments.

For the last decades, the progress in theoretical particle physics has
been driven by naturalness considerations in the form of the famous
hierarchy problem. This has led to many extensions of the SM, all of
them attempting to explain why the weak scale has not been pushed to
much higher energy scales by some hypothetical NP degrees of
freedom. Among the many proposals to address this issue, supersymmetry
(SUSY) is certainly the most popular one. However, and similarly to
other analogous solutions to the hierarchy problem, the predicted new
particles at the weak scale have not been observed at the LHC.

This has of course raised some doubts about the existence of
supersymmetry close to the weak scale. Since this proximity is to be
expected in case supersymmetry has something to do with the hierarchy
problem, the whole idea of weak scale supersymmetry is under some
pressure at the moment. However, it is worth keeping in mind that most
experimental searches for SUSY focus on the Minimal Supersymmetric
Standard Model (MSSM). This model, which constitutes the minimal
extension of the SM that incorporates SUSY, has some underlying
assumptions that lead to very specific signatures. For example, in the
MSSM one assumes the conservation of a discrete symmetry, known as
R-parity \cite{Fayet:1974pd,Farrar:1978xj}, which forbids all
renormalizable lepton and baryon number violating operators and leads
to the existence of a stable particle, the lightest supersymmetric
particle (LSP), which in turn leads to large amounts of missing
energy in supersymmetric events at the LHC. There are, however, many
known (and well motivated) supersymmetric scenarios with R-parity
violation and, in fact, several authors have shown that simply by
allowing for non-zero B-violating terms in the superpotential, the
current LHC bounds can be clearly relaxed, allowing for the existence
of light squarks and gluinos hidden in the huge QCD
background~\cite{Evans:2012bf,Bhattacherjee:2013gr}. Similarly, one
can extend the MSSM in many other directions, often changing the
phenomenology at colliders dramatically. This suggests that it might
be too soon to give up on SUSY, a framework with many possibilities
yet to be fully explored.

As explained above, there are some well-grounded phenomenological
issues that cannot be explained within the SM. One of these open
problems is the existence of non-zero neutrino masses and mixings,
nowadays firmly by neutrino oscillation
experiments~\cite{Forero:2014bxa,Gonzalez-Garcia:2014bfa,Capozzi:2013csa}. In
fact, this issue is not addressed in the MSSM either, since neutrinos
remain massless in the same way as in the SM. This calls for an
extension of the MSSM that extends the lepton sector and accommodates
the observations in neutrino oscillation experiments. This can be done
in two different ways: (1) \textit{high-energy extensions}, in which
the new degrees of freedom responsible for the generation of neutrino
masses live at very high energy scales, and (2) \textit{low-energy
  extensions}, with new particles and/or interactions at the SUSY
scale.

One of the most generic predictions in neutrino mass models is lepton
flavor violation (LFV). In fact, neutrino oscillations are the proof
that lepton flavor is not a conserved symmetry of nature, since
neutrinos produced with a given flavor change it as they
propagate. Therefore, all neutrino mass models built to give an
explanation to oscillation experiments violate lepton flavor. However,
we have never observed LFV processes involving charged leptons
although, in principle, there is no symmetry (besides lepton flavor,
which we know to be broken) that forbids processes like $\mu^- \to e^-
\gamma$, $\tau^- \to e^- \mu^+ \mu^-$ or $K_L \to e^- \mu^+$. This
fact can be well understood in some minimal frameworks, such as the
minimal extension of the SM with Dirac neutrinos. In this case, LFV in
the charged lepton sector is strongly suppressed, since neutrino
masses are the only source of LFV, leading to unobservable LFV rates,
like $\text{BR}(\mu^- \to e^- \gamma) \sim
10^{-55}$~\cite{Petcov:1976ff}. However, as soon as one extends the
SM, this conclusion can be clearly
altered~\cite{Cheng:1976uq,Bilenky:1977du}. In fact, new sources of
LFV can be found in most extensions of the leptonic sector, either
caused by new interactions, by new particles or even by complete new
sectors that couple to the SM leptons.

After this discussion on LFV and neutrino masses a clarification is in
order. Although neutrino oscillations imply LFV, LFV does not
necessarily imply neutrino oscillations. There are models that predict
charged lepton LFV without generating a mass for the neutrinos. The
simplest example of this class of models is the general
Two-Higgs-Doublet of type-III, where neutrinos remain massless but
lepton flavor is violated due to the existence of off-diagonal
$h-\ell_i-\ell_j$ vertices. Another relevant example is the MSSM
itself, where neutrinos are also massless, but the slepton soft masses
can induce LFV processes if they contain off-diagonal entries. One can
actually estimate the branching ratio for the radiative LFV decay
$\ell_i \to \ell_j \gamma$ as \cite{Kuno:1999jp}
\begin{equation} \label{eq:BRestimate}
 \text{BR}(\ell_i \to \ell_j \gamma) \simeq \frac{48 \pi^3 \alpha}{G_F^2}
\frac{|(m^2_{\tilde \ell})_{ij}|^2}{M^8_{SUSY}} \text{BR}(\ell_i \to \ell_j \nu_i{\bar\nu_j}) \, ,
\end{equation}
where $G_F$ is the Fermi constant, $\alpha$ the fine structure
constant, $(m^2_{\tilde \ell})_{ij}$ are the dominant off-diagonal
elements of the soft SUSY breaking slepton mass matrices and
$M_{SUSY}$ is the typical mass of the SUSY particles, expected to be
in the TeV ballpark. This estimate clearly shows that rather small
off-diagonal elements are required to satisfy the experimental bounds
\cite{Agashe:2014kda}.

In general, one concludes that large LFV rates are expected in most
models beyond the SM. This motivates the study of LFV as an indirect
probe of new physics and, in particular, of supersymmetric models
beyond the MSSM. This is the subject of this review~\footnote{The
  field of lepton flavor violation beyond the MSSM has been intensely
  explored for many years and contains a vast literature. In this
  review I present my personal view of the subject and thus I must
  apologize for those papers which are not cited.}. In particular, we
will concentrate on three different scenarios: high-scale and
low-scale seesaw models as well as models with R-parity violation. As
we will see, the LFV phenomenology turns out to be very different
depending on the exact scenario, implying that lepton flavor violation
may be richer than in the MSSM. In some cases the common lore
(established in the MSSM) turns out to be wrong, and specific studies
must be performed in order to correctly understand the corresponding
LFV phenomenology.

This review is organized as follows: in Sec.~\ref{sec:LFV} we give an
overview of the current experimental situation and briefly discuss
some projects that will take place in the near future. Then we review
the LFV phenomenology of three different types of models beyond the
MSSM: high-scale seesaw models (in Sec. \ref{sec:high}), low-scale
seesaw models (in Sec. \ref{sec:low}) and models with R-parity
violation (in Sec. \ref{sec:rpv}). Finally, we conclude in
Sec.~\ref{sec:conclusions}.

\section{Current experimental situation and future projects} \label{sec:LFV}

The search for LFV is soon going to live a \textit{golden age} given
the upcoming experiments devoted to high-intensity
physics~\footnote{See
  \cite{Bernstein:2013hba,Mihara:2013zna,Signorelli:2013kla} for
  recent reviews.}. In addition to the LFV searches already taking
place in several experiments, new projects will join the effort in the
next few years.

In what concerns the radiative decay $\ell_i \to \ell_j \gamma$, the
experiment leading to the most stringent constraints is MEG. This
experiment, located at the Paul Scherrer Institute in Switzerland,
searches for the radiative process $\mu \to e \gamma$. Recently, the
MEG collaboration announced a new limit on the rate for this process
based on the analysis of a dataset with $3.6 \times 10^{14}$ stopped
muons. The non-observation of the LFV process led to the limit
$\text{BR}(\mu \to e \gamma) < 5.7 \cdot 10^{-13}$
\cite{Adam:2013mnn}, four times more stringent than the previous limit
obtained by the same collaboration. Moreover, the MEG collaboration
has announced plans for future upgrades. These will allow to reach a
sensitivity of about $6 \cdot 10^{-14}$ after 3 years of acquisition
time \cite{Baldini:2013ke}. This is of great
importance, as this observable is the one with the largest rates in
many models.

The most promising improvements in the near future are expected in
$\mu \to 3 \, e$ and $\mu-e$ conversion in nuclei. Regarding the
former, the decay $\mu \to 3 \, e$ was searched for long ago by the
SINDRUM experiment \cite{Bellgardt:1987du}, setting the strong limit
$\text{BR}(\mu \to 3 \, e) < 1.0 \cdot 10^{-12}$. The future Mu3e
experiment announces a sensitivity of $\sim 10^{-16}$
\cite{Blondel:2013ia}, which would imply an impressive improvement by
$4$ orders of magnitude. As for the latter, several experiments will
compete in the next few years, with sensitivities for the conversion
rate ranging from $10^{-14}$ to an impressive $10^{-18}$. These
include Mu2e \cite{Carey:2008zz,Glenzinski:2010zz,Abrams:2012er},
DeeMe \cite{Aoki:2010zz}, COMET \cite{Cui:2009zz,Kuno:2013mha} and the
future PRISM/PRIME \cite{Barlow:2011zza}. In all cases, these experiments will
definitely improve on previous experimental limits.

The limits for $\tau$ observables are less stringent, although
significant improvements are expected from B factories like Belle II
\cite{Aushev:2010bq,Bevan:2014iga}. Finally, although the most common
way to search for LFV is in low-energy experiments, colliders can also
play a very relevant role looking for LFV processes at high
energies. The LHCb collaboration reported recently the first bounds on
$\tau\to 3 \, \mu$ ever obtained in a hadron collider
\cite{Aaij:2013fia}. Furthermore, the CMS collaboration recently found
an intriguing $2.4 \sigma$ excess in the $h\to \tau \mu$ channel which
translates into $\text{BR}(h\to\tau\mu) = \left( 0.84_{-0.37}^{+0.39}
\right)$\%~\cite{Khachatryan:2015kon}. For reference, in
Tab.~\ref{tab:sensi} we collect present bounds and expected
near-future sensitivities for the most popular low-energy LFV
observables.

\begin{table}[tb!]
\centering
\begin{tabular}{|c|c|c|}
\hline
LFV Process & Present Bound & Future Sensitivity  \\
\hline
    $\mu \rightarrow  e \gamma$ & $5.7\times 10^{-13}$~\cite{Adam:2013mnn}  & $6\times 10^{-14}$~\cite{Baldini:2013ke} \\
    $\tau \to e \gamma$ & $3.3 \times 10^{-8}$~\cite{Aubert:2009ag}& $ \sim3\times10^{-9}$~\cite{Aushev:2010bq}\\
    $\tau \to \mu \gamma$ & $4.4 \times 10^{-8}$~\cite{Aubert:2009ag}& $ \sim3\times10^{-9}$~\cite{Aushev:2010bq} \\
    $\mu \rightarrow e e e$ &  $1.0 \times 10^{-12}$~\cite{Bellgardt:1987du} &  $\sim10^{-16}$~\cite{Blondel:2013ia} \\
    $\tau \rightarrow \mu \mu \mu$ & $2.1\times10^{-8}$~\cite{Hayasaka:2010np} & $\sim 10^{-9}$~\cite{Aushev:2010bq} \\
    $\tau^- \rightarrow e^- \mu^+ \mu^-$ &  $2.7\times10^{-8}$~\cite{Hayasaka:2010np} & $\sim 10^{-9}$~\cite{Aushev:2010bq} \\
    $\tau^- \rightarrow \mu^- e^+ e^-$ &  $1.8\times10^{-8}$~\cite{Hayasaka:2010np} & $\sim 10^{-9}$~\cite{Aushev:2010bq} \\
    $\tau \rightarrow e e e$ & $2.7\times10^{-8}$~\cite{Hayasaka:2010np} &  $\sim 10^{-9}$~\cite{Aushev:2010bq} \\
    $\mu^-, \mathrm{Ti} \rightarrow e^-, \mathrm{Ti}$ &  $4.3\times 10^{-12}$~\cite{Dohmen:1993mp} & $\sim10^{-18}$~\cite{PRIME} \\
    $\mu^-, \mathrm{Au} \rightarrow e^-, \mathrm{Au}$ & $7\times 10^{-13}$~\cite{Bertl:2006up} & \\
    $\mu^-, \mathrm{Al} \rightarrow e^-, \mathrm{Al}$ &  & $10^{-15}-10^{-18}$ \\
    $\mu^-, \mathrm{SiC} \rightarrow e^-, \mathrm{SiC}$ &  & $10^{-14}$~\cite{Natori:2014yba} \\
\hline
\end{tabular}
\caption{Current experimental bounds and future sensitivities for the most important LFV observables.}
\label{tab:sensi}
\end{table}

The theoretical understanding of all these processes will be crucial
in case a discovery is made. With such a large variety of processes,
the determination of hierarchies or correlations in specific models
will allow us to extract fundamental information on the underlying
physics behind LFV. This goal requires detailed analytical and
numerical studies of the different contributions to the LFV processes,
in order to get a global picture of the LFV \textit{anatomy} of the
relevant models and be able to discriminate among them by means of
combinations of observables with definite predictions
\cite{Buras:2010cp}.

\section{High-scale seesaw models} \label{sec:high}

Neutrino mixing is, by itself, a flavor violating effect. Therefore,
all neutrino mass models that aim at explaining the observed pattern
of neutrino masses and mixings incorporate lepton flavor
violation. However, specific predictions can be very different in
different models.

Among the huge number of scenarios proposed for neutrino mass
generation, the seesaw mechanism is arguably the most popular one. In
its conventional form, the seesaw mechanism explains the smallness of
neutrino mass by means of a very large energy scale, the
\textit{seesaw scale} $M_{SS}$, which suppresses neutrino masses as
\begin{equation} \label{eq:seesaw}
m_\nu \sim \frac{v^2}{M_{SS}} \, .
\end{equation}
Here $\langle H^0 \rangle = v/\sqrt{2} = 174$ GeV is the standard
Higgs boson vacuum expectation value (VEV) that determines the weak
scale. In order to obtain neutrino masses of about $\sim 0.1$ eV, one
requires $M_{SS} \sim 10^{14}$ GeV. For this reason, this setup is
usually called \textit{high-scale seesaw}. The proximity of the
high-energy scale $M_{SS}$ to the grand unification (GUT) scale (as
predicted in the MSSM) $m_{\text{GUT}} = 2 \cdot 10^{16}$ GeV suggests an
intriguing connection with unification physics, making the seesaw a
very well-motivated scenario.

Regarding specific realizations of the seesaw mechanism, it is
well-known that with renormalizable interactions only, three
tree-level realizations exist~\cite{Ma:1998dn}. These are usually
called
type-I~\cite{Minkowski:1977sc,Yanagida:1979,GellMann:1980vs,Mohapatra:1979ia,Schechter:1980gr,Schechter:1981cv},
II~\cite{Konetschny:1977bn,Schechter:1980gr,Marshak:1980yc,Lazarides:1980nt,Mohapatra:1980yp,
  Cheng:1980qt,Schechter:1981cv} and III~\cite{Foot:1988aq}. They
differ from each other by the nature of the seesaw messengers: in the
type-I seesaw these are singlet right-handed neutrinos, in the type-II
seesaw scalar $SU(2)_L$ triplets with hypercharge two, and in the
type-III seesaw fermionic $SU(2)_L$ triplets with vanishing
hypercharge. In all cases they lead to a neutrino mass of the form of
Eq. \eqref{eq:seesaw}, where $M_{SS}$ is proportional to the mass of
the heavy mediators, and the induced neutrino masses are of Majorana
type, thus breaking lepton number in two units.

Given the large Majorana masses of the seesaw mediators, one may
wonder about how to probe the high-scale seesaw. In supersymmetric
scenarios this is possible thanks to the sleptons. Even if their soft
terms are flavor conserving at some high-energy scale, the
renormalization group running down to the SUSY scale will induce
non-zero off-diagonal terms due to their interactions with the seesaw
mediators~\cite{Borzumati:1986qx}. These can be probed since the
misalignment of the slepton mass matrices with respect to that of the
SM charged leptons induces LFV processes such as $\ell_i \to \ell_j
\gamma$, $\ell_i \to 3 \, \ell_j$ and $\mu-e$ conversion in
nuclei. This connection between the phenomenology at low-energies and
the high-scale mediators is only possible in supersymmetric models and
constitutes an excelent opportunity to test the standard seesaw
scenario~\footnote{In the non-SUSY version of the seesaw mechanism
  this link between high and low energy scales is lost. In this case
  probing the origin of neutrino masses becomes a quite challenging
  task, and only very indirect probes such as neutrinoless double beta
  decay are possible \cite{Deppisch:2012nb}.}.

In case of high-scale seesaw models, the low-energy theory is simply
the MSSM. This allows one to establish definite patterns and
hierarchies among the LFV observables. For instance, the branching
ratios for the $\ell_i \to \ell_j \gamma$ and $\ell_i \to 3 \, \ell_j$
LFV decays follow the approximate relation
\cite{Ilakovac:1994kj,Hisano:1995cp,Arganda:2005ji},
\begin{equation} \label{eq:dipole}
\text{BR}(\ell_i \to 3 \, \ell_j) \simeq \frac{\alpha}{3 \pi} 
\left(\log\left(\frac{m^2_{\ell_i}}{m^2_{\ell_j}}\right) - \frac{11}{4} \right) 
\text{BR}(\ell_i \to \ell_j \gamma) \, .
\end{equation}
Therefore, in supersymmetric high-scale seesaw models, the most
constraining LFV process is $\ell_i \to \ell_j \gamma$. The relation
in Eq. \eqref{eq:dipole} is caused by the so-called \emph{dipole
  dominance} in high-scale seesaw models. Among the different
contributions to the 3-body decay $\ell_i \to 3 \, \ell_j$, the dipole
photon penguins are the dominant ones, leading to the proportionality
between the $\ell_i \to \ell_j \gamma$ and $\ell_i \to 3 \, \ell_j$
branching ratios~\footnote{An exception to this general rule is found
  for low pseudoscalar masses and large $\tan \beta$
  \cite{Babu:2002et}. In this case, Higgs penguins turn out to be
  dominant in processes involving the second and third generations,
  like $\tau \to 3 \, \mu$. However, this region of parameter space is
  nowadays under some tension due to strong flavor constraints derived
  from the observation of quark flavor violating processes like $B_s
  \to \mu^+ \mu^-$ \cite{Aaij:2012nna}.}.

\subsection{Standard high-scale seesaw scenarios} \label{subsec:minimalseesaw}

Implementing a high-scale seesaw mechanism in supersymmetric scenarios
involves an additional complication. This is related to one of the
most appealing features of the MSSM: gauge coupling unification. In
case of the type-I seesaw, the introduction of the seesaw mediator
does not spoil this attractive feature, since the right-handed
neutrino superfields are gauge singlets and do not affect the running of
the gauge couplings. In contrast, in the type-II and type-III seesaws,
new contributions to the running of the $SU(2)_L$ and $U(1)_Y$ gauge
couplings are induced by the seesaw mediators. However, a well-known
solution to this problem exists. Unification can be easily restored by
embedding the seesaw mediators in full $SU(5)$ multiplets, like {\bf
  15}-plets in the case of type-II~\cite{Rossi:2002zb} or {\bf
  24}-plets~\cite{Buckley:2006nv} in the case of type-III. The
contributions from the other members of the multiplet guarantee that
the three gauge couplings will eventually meet at a high energy scale,
$m_{\text{GUT}}$, although the common value of the coupling changes,
$g_{\text{GUT}}$, might be different from that of the MSSM. In addition, note
that the {\bf 24}-plet of $SU(5)$ contains, besides the $SU(2)_L$
triplet, a singlet state which also contributes to neutrino
masses. Hence, in this case one actually has a mixture between type-I
and type-III seesaws.

The new superfield content, explicitly denoting gauge charges under
$SU(3)_c \times SU(2)_L \times U(1)_Y$, and superpotential for each
seesaw variant are~\cite{Hirsch:2012ti}:
\begin{itemize}

\item {\bf Type-I:} Three generations of right-handed neutrino
  superfields, singlets of $SU(5)$, are introduced, $\widehat N^c \sim
  (1,1,0)$.

\begin{equation} \label{eq:typeI}
W_{I} = W_\mathrm{MSSM} + Y_\nu \widehat N^c \widehat L \widehat H_u + \frac{1}{2} M_R \widehat N^c \widehat N^c
\end{equation}

\item {\bf Type-II:} In this case one needs to introduce a vector-like
  pair of {\bf 15} and $\boldsymbol{\overline{15}}$ of $SU(5)$, decomposed as
  $\widehat S \sim (6,1,-2/3)$, $\widehat T \sim (1,3,1)$ and
  $\widehat Z \sim (3,2,1/6)$ (as well as the corresponding bar
  superfields). $\widehat T$ and $\widehat{\overline{T}}$ are the
  $SU(2)_L$ triplets responsible for neutrino mass generation.

\begin{eqnarray}
W_{II} & = & W_\mathrm{MSSM} + \frac{1}{\sqrt{2}}(Y_T \widehat L \widehat T  \widehat L
+  Y_S \widehat d^c \widehat S \widehat d^c)
+ Y_Z \widehat d^c \widehat Z \widehat L   \nonumber \\
& + & \frac{1}{\sqrt{2}}(\lambda_1 \widehat H_d \widehat T \widehat H_d
+\lambda_2  \widehat H_u \widehat{\overline{T}} \widehat H_u)
+ M_T \widehat T \widehat{\overline{T}}
+ M_Z \widehat Z \widehat{\overline{Z}} + M_S \widehat S \widehat{\overline{S}} \label{eq:typeII}
\end{eqnarray}

\item {\bf Type-III:} Three generations of {\bf 24} of $SU(5)$ are
  added. They can be decomposed as $\widehat N^c \sim (1,1,0)$,
  $\widehat G \sim (8,1,0)$, $\widehat \Sigma \sim (1,3,0)$, $\widehat
  X \sim (3,2,-5/6)$ and $\widehat{\overline{X}} \sim (\bar 3,2,5/6)$. As
  explained above, neutrino masses are generated as a combination of a
  type-I seesaw (mediated by $N^c$) and a type-III seesaw (mediate by
  the $SU(2)_L$ triplet $\Sigma$).

\begin{eqnarray}
 W_{III} & = &  W_\mathrm{MSSM}
 +  \widehat H_u\left( Y_\Sigma \widehat \Sigma - \sqrt{\frac{3}{10}}
               Y_\nu \widehat N^c \right) \widehat L
 + Y_X \widehat H_u \widehat{\overline{X}} \widehat d^c \nonumber \\
         & & + \frac{1}{2} M_{R} \widehat N^c \widehat N^c
         + \frac{1}{2} M_{G} \widehat G \widehat G
          + \frac{1}{2} M_{\Sigma} \widehat \Sigma \widehat \Sigma
          + M_{X} \widehat X \widehat{\overline{X}}  \label{eq:typeIII}
\end{eqnarray}

\end{itemize}

The following notation is used in Eqs. \eqref{eq:typeI},
\eqref{eq:typeII} and \eqref{eq:typeIII}: $W_\mathrm{MSSM}$ is the MSSM
superpotential, $\widehat H_d$, $\widehat H_u$ and $\widehat L$ are
the down-Higgs, up-Higgs and lepton $SU(2)_L$ doublet superfields,
respectively, and $\widehat d^c$ is the right-handed down-type quark
superfield.

The LFV phenomenology of SUSY seesaw models has been studied by many
authors. For the type-I seesaw, low-energy LFV decays such as $\ell_i
\to \ell_j \gamma$ and $\ell_i \to 3 \, \ell_j$ have been calculated
in
\cite{Hisano:1995nq,Hisano:1995cp,Ellis:2002fe,Deppisch:2002vz,Petcov:2003zb,
  Arganda:2005ji,Petcov:2005yh,Antusch:2006vw,Deppisch:2004fa,Hirsch:2008dy,Abada:2010kj,Abada:2012re,Figueiredo:2013tea}.
Similarly, $\mu-e$ conversion in nuclei has been studied in
\cite{Arganda:2007jw,Deppisch:2005zm}. The other two seesaw variants
have received much less attention. The LFV phenomenology of the SUSY
type-II seesaw has been considered in
\cite{Rossi:2002zb,Joaquim:2006uz,Joaquim:2006mn,Hirsch:2008gh,Esteves:2009qr,Joaquim:2009vp,Brignole:2010nh},
whereas the SUSY type-III seesaw has been studied in
\cite{Esteves:2010ff,Abada:2011mg,Hirsch:2012yv}. More recently, the
interplay between the Higgs mass constraint and LFV was studied in
\cite{Hirsch:2012ti} for the three seesaw variants. In the following
we comment on some selected results.

\begin{figure}[t!]
\centering
\includegraphics[width=0.7 \linewidth]{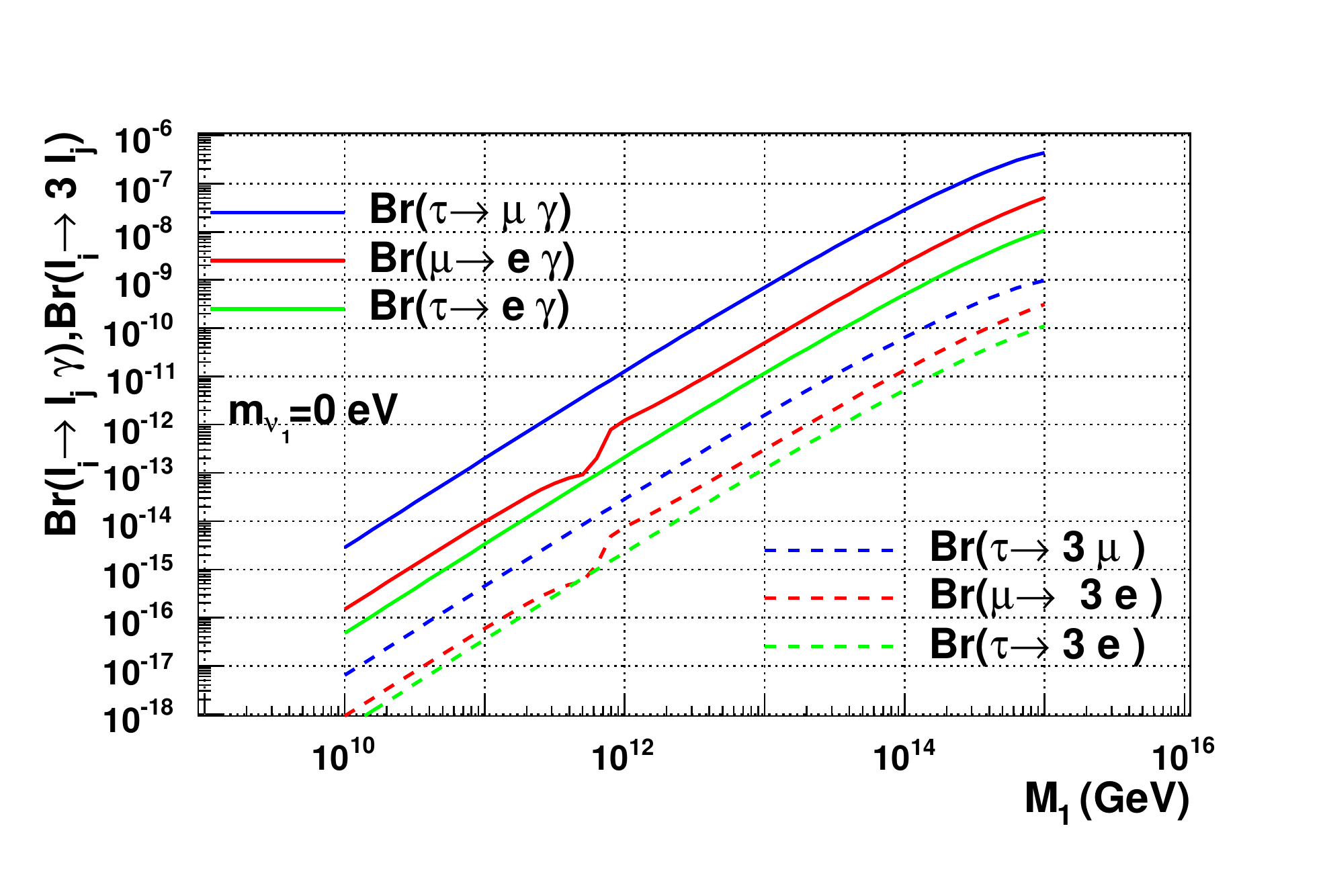}
\caption{Branching ratios for $\ell_i \to \ell_j \gamma$ and $\ell_i
  \to 3 \, \ell_j$ as a function of the seesaw scale in the SUSY
  type-I seesaw. This figure was obtained in the standard SPS1a'
  point, assuming degenerate right-handed neutrinos and fixing the
  neutrino Yukawas to reproduce tribimaximal mixing. Furthermore, a
  massless lightest neutrino was also assumed. Figure taken from
  \cite{Hirsch:2008dy}.}
\label{fig:typeI}
\end{figure}

Let us first comment on some results for the SUSY type-I seesaw.
Fig. \ref{fig:typeI} shows the ranching ratios for the $\ell_i \to
\ell_j \gamma$ and $\ell_i \to 3 \, \ell_j$ decays as a function of
the seesaw scale (the mass of the right-handed neutrino mass). This
figure was obtained in Ref. \cite{Hirsch:2008dy}, using the standard
SPS1a' point \cite{Allanach:2002nj}, assuming degenerate right-handed
neutrinos and a massless lightest neutrino and fixing the neutrino
Yukawas to reproduce tribimaximal mixing. Although this parameter
choice is nowadays excluded for several reasons (the SUSY spectrum is
too light to pass the constraints from LHC searches and tribimaximal
mixing is now excluded after $\theta_{13}$ has been measured), it
serves to illustrate the dipole dominance discussed above. Indeed, one
sees a perfect correlation between the branching ratios of $\ell_i \to
\ell_j \gamma$ and $\ell_i \to 3 \, \ell_j$, with $\text{BR}(\ell_i
\to 3 \, \ell_j) \ll \text{BR}(\ell_i \to 3 \, \ell_j)$. As already
discussed, this is due to the fact that the photonic dipole operator
($\bar \ell_i F_{\mu \nu} \sigma^{\mu \nu} \ell_j$) dominates both
processes.

We now turn ot the SUSY type-II seesaw. In the type-II seesaw, the
neutrino mass matrix is proportional to the $Y_T$ Yukawa matrix,
\begin{equation} \label{eq:typeIImass}
m_\nu=\frac{v_u^2}{2} \frac{\lambda_2}{M_T}Y_T \, .
\end{equation}
This is derived from the superpotential term $Y_T \widehat L \widehat
T \widehat L$ in Eq. \eqref{eq:typeII}. This direct relation has
important consequences for the phenomenology, since it forces the
flavor structure of $Y_T$ to be the same as that of $m_\nu$, the
latter being \textit{measured} in neutrino oscillation
experiments~\footnote{In contrast, in the type-I and type-III seesaws
  the analogous relation is quadratic in the Yukawa coupling. This
  introduces extra freedom in the determination of the seesaw
  parameters (usually encoded in the so-called $R$ matrix
  \cite{Casas:2001sr}) and makes it impossible to predict the Yukawa
  flavor structure only from neutrino oscillation data.}. In other
words, if all the neutrino masses, angles and phases were known, $Y_T$
would be completely fixed (up to an overall constant). Since $Y_T$
determines the LFV phenomenology, this implies correlations between
the neutrino oscillation parameters and LFV observables.

\begin{figure}[t!]
\centering
\includegraphics[width=0.48 \linewidth]{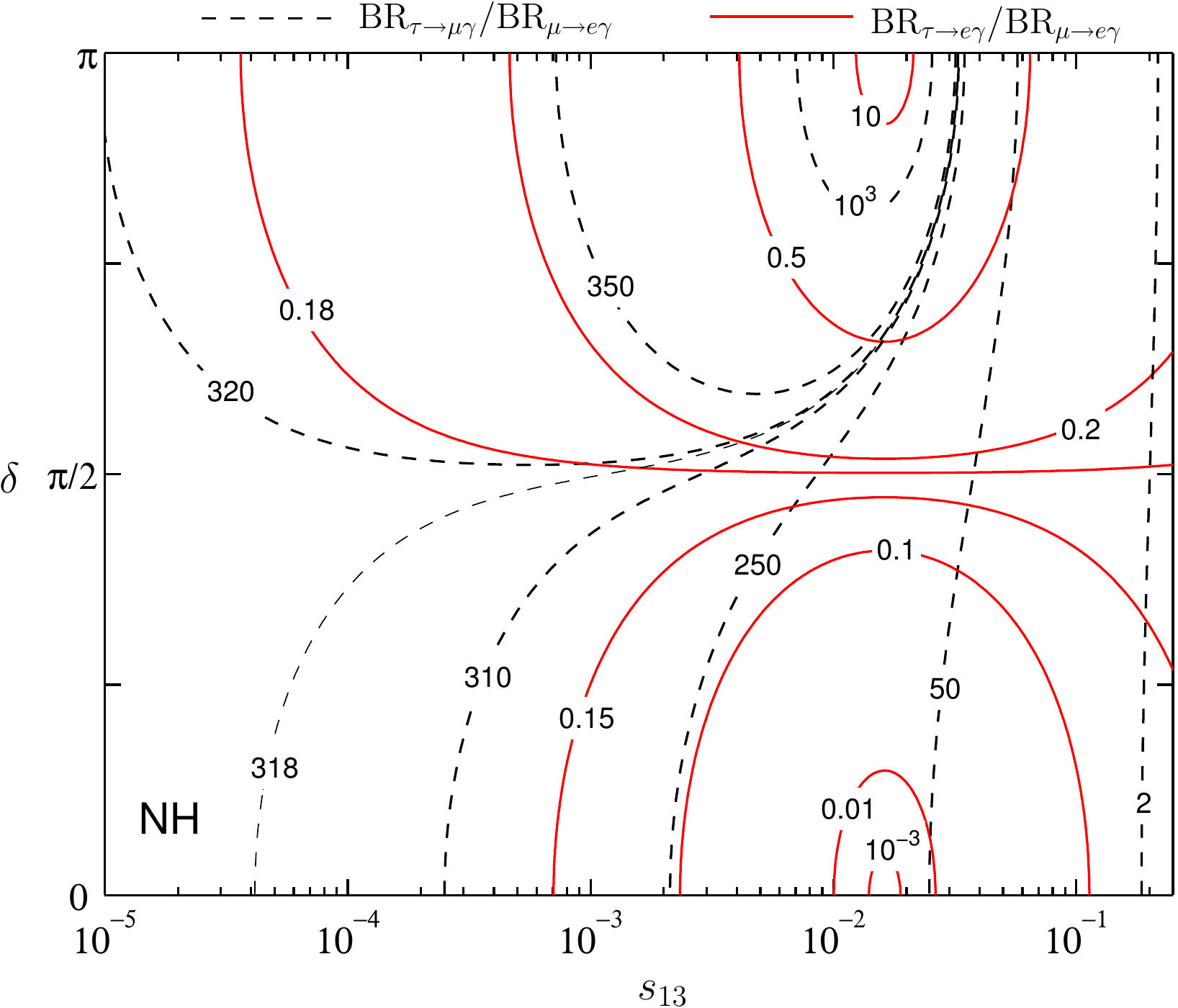}
\includegraphics[width=0.48 \linewidth]{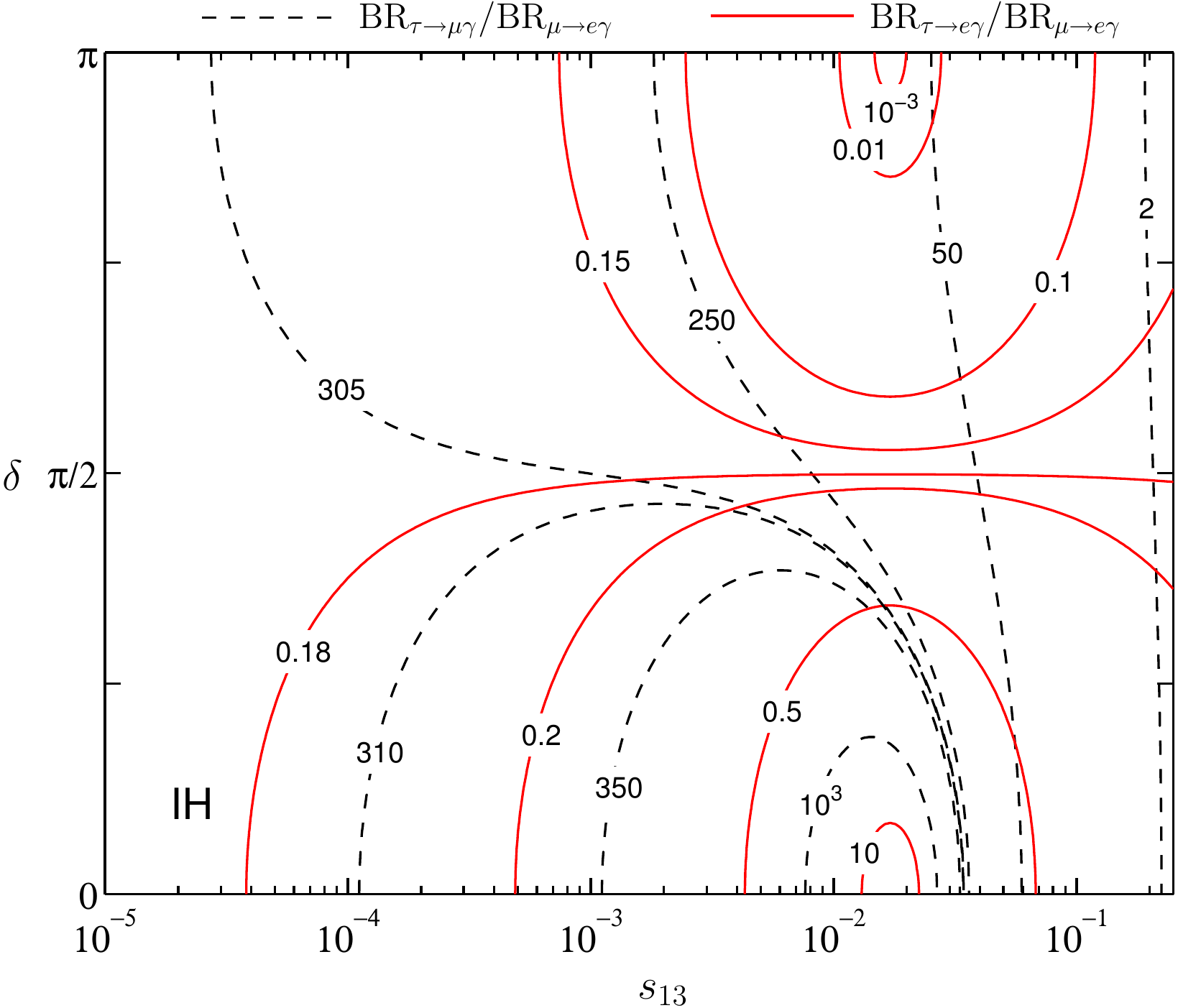}
\caption{Contours of the ratios $\text{BR}(\tau \to \mu
  \gamma)/\text{BR}(\mu \to e \gamma)$ (black, dashed lines) and
  $\text{BR}(\tau \to e \gamma)/\text{BR}(\mu \to e \gamma)$ (red,
  solid lines) in the $(\sin \theta_{13} , \delta)$ plane, for normal
  hierarchy (on the left) and inverted hierarchy (on the right) for
  the neutrino mass spectrum. Figure taken from
  \cite{Joaquim:2006mn}.}
\label{fig:typeII}
\end{figure}

A clear illustration of the previous point is shown in
Fig. \ref{fig:typeII}, borrowed from Ref. \cite{Joaquim:2006mn}. By
computing the ratios $\text{BR}(\ell_i \to \ell_j
\gamma)/\text{BR}(\ell_m \to \ell_n \gamma)$ one gets rid of the
unknown overall factor in the $Y_T$ Yukawas, thus obtaining direct
predictions in terms of neutrino parameters. In this case, the figure
shows the dependence of these ratios on the mixing angle $\theta_{13}$
and the Dirac CP violating phase $\delta$. We see that this scenario
is extremely predictive. For example, finding experimentally
$\text{BR}(\tau \to e \gamma) > \text{BR}(\mu \to e \gamma)$ would
immediately rule out the model, at least in its minimal
form~\footnote{One way to spoil these strict predictions is to
  introduce a second $SU(2)_L$ triplet $T^\prime$. In this case
  $m_\nu$ would receive contributions from $T$ and $T^\prime$, $m_\nu
  = m_\nu^T + m_\nu^{T^\prime}$, and the proportionality in
  Eq. \eqref{eq:typeIImass} would be lost.}.

Additional ways to test high-scale SUSY seesaws include slepton mass
splittings \cite{Buras:2009sg} (directly related to LFV) and the study
of the SUSY spectrum, usually \textit{deformed} with respect to the
standard spectra in constrained (CMSSM) scenarios. In particular, one
can construct certain \textit{invariants} that contain information
about the high-energy scale, see for example
\cite{Buckley:2006nv,Hirsch:2011cw,Arbelaez:2011bb}. See also
\cite{Biggio:2012wx} for related ideas.

\subsection{Extended high-scale seesaw scenarios} \label{subsec:extendedseesaw}

We now turn our attention to extended high-scale SUSY seesaw scenarios
beyond the \textit{classical} type-I, type-II and type-III
seesaws. However, before we concentrate on the extended models, let us
make a general observation. As already discussed, flavor violating
entries in the slepton soft terms $m_{\tilde L}^2$ and $m_{\tilde
  e}^2$ (the left and right slepton squared soft masses, respectively)
are induced due to their interactions with the seesaw mediators. Even
if they are flavor diagonal at the unification scale, off-diagonal
terms are generated at low energies by renormalization group running,
thus inducing all kinds of LFV processes. In the case of the radiative
$\ell_i \to \ell_j \gamma$, the effective dipole operator that
contributes to the decay can be written as
\begin{equation}
\mathcal{L}_{\text{dipole}} = e \, \frac{m_{\ell_i}}{2} \, \bar{\ell}_i \sigma_{\mu \nu} F^{\mu \nu} 
(A_L^{ij} P_L + A_R^{ij} P_R) \ell_j + \, \text{h.c.} \, ,
\end{equation}
where $P_{L,R} = \frac{1}{2}(1 \mp \gamma_5)$ are the usual chirality
projectors and $e$ the electric charge. The Wilson coefficients $A_L$
and $A_R$ are generated by loops with left and right sleptons,
respectively. One finds
\begin{equation} \label{A-dependence}
A_L^{ij} \sim \frac{(m_{\tilde L}^2)_{ij}}{M_{SUSY}^4} \quad , \quad A_R^{ij} 
\sim \frac{(m_{\tilde e}^2)_{ij}}{M_{SUSY}^4} \, ,
\end{equation}
where it has been assumed that A-terms mixing left-right transitions
are negligible. $\text{BR}(\ell_i \to \ell_j \gamma)$ can be computed
in terms of $A_L$ and $A_R$ as
\begin{equation} \label{eq:BRestimate2}
 \text{BR}(\ell_i \to \ell_j \gamma) \simeq \frac{48 \pi^3 \alpha}{G_F^2}
\left( \left| A_L^{ij} \right|^2 + \left| A_R^{ij} \right|^2 \right) \text{BR}(\ell_i \to \ell_j \nu_i{\bar\nu_j}) \, ,
\end{equation}
The straightforward combination of Eqs. \eqref{A-dependence} and
\eqref{eq:BRestimate2} leads to Eq. \eqref{eq:BRestimate}.

In the minimal SUSY seesaw models discussed above, the seesaw
mediators only couple to the left sleptons. For instance, in the
type-I case this interaction is given by the superpotential coupling
$Y_\nu \widehat L \widehat H_u \widehat N$, whereas in the type-II
case it is given by the $Y_T \widehat L \widehat T \widehat L$
term. For this reason, negligible off-diagonal entries in $m_{\tilde
  e}^2$ are induced, implying that minimal SUSY seesaw models predict
$A_R \simeq 0$. As we will see below, this has an impact on some
low-energy observables that allow, in principle, to test the
minimality of the high-scale seesaw mechanism.

\subsubsection*{Supersymmetric models with non-minimal seesaw mechanisms}

As an example supersymmetric model with a non-minimal seesaw
mechanisms, we consider the left-right symmetric model of
\cite{Aulakh:1997ba,Aulakh:1997fq} (in the following simply called
`the LR model'). The LFV and dark matter phenomenology of this model
has been studied in detail in \cite{Esteves:2010si,Esteves:2011gk}.

The model is defined below the GUT scale~\footnote{The model
  implicitly assumes the existence of a GUT model at higher
  energies. At $m_{\text{GUT}}$, the gauge couplings and soft terms unify.},
where the gauge group is $SU(3)_c \times SU(2)_L \times SU(2)_R \times
U(1)_{B-L}$. In addition, we assume that parity is conserved. The
matter content of the model is given in
Tab. \ref{tab:particles-step1}. Here $\widehat Q$, $\widehat Q^c$,
$\widehat L$ and $\widehat L^c$ are the quark and lepton superfields
of the MSSM with the addition of (three) right-handed neutrino
superfields to complete the $\widehat L^c$ $SU(2)_R$ doublets.

Two $\widehat \Phi$ superfields, bidoublets under $SU(2)_L \times
SU(2)_R$, are introduced. Among their components, they contain the
standard $\widehat H_d$ and $\widehat H_u$ MSSM Higgs doublets.
Finally, the rest of the superfields in Tab.
\ref{tab:particles-step1} are introduced to break the LR symmetry.

\begin{table}
\centering
\begin{tabular}{c c c c c c}
\hline
Superfield & generations & $SU(3)_c$ & $SU(2)_L$ & $SU(2)_R$ & $U(1)_{B-L}$ \\
\hline
$\widehat Q$ & 3 & 3 & 2 & 1 & $\frac{1}{3}$ \\
$\widehat Q^c$ & 3 & $\bar{3}$ & 1 & 2 & $-\frac{1}{3}$ \\
$\widehat L$ & 3 & 1 & 2 & 1 & -1 \\
$\widehat L^c$ & 3 & 1 & 1 & 2 & 1 \\
$\widehat \Phi$ & 2 & 1 & 2 & 2 & 0 \\
$\widehat \Delta$ & 1 & 1 & 3 & 1 & 2 \\
$\widehat {\bar \Delta}$ & 1 & 1 & 3 & 1 & -2 \\
$\widehat \Delta^c$ & 1 & 1 & 1 & 3 & -2 \\
$\widehat {\bar \Delta}^c$ & 1 & 1 & 1 & 3 & 2 \\
$\widehat \Omega$ & 1 & 1 & 3 & 1 & 0 \\
$\widehat \Omega^c$ & 1 & 1 & 1 & 3 & 0 \\
\hline
\end{tabular}
\caption{LR model. Matter content between the GUT scale and the
  $SU(2)_R$ breaking scale. The electric charge operator is defined as
  $Q = I_{3L} + I_{3R} + \frac{B-L}{2}$.}
\label{tab:particles-step1}
\end{table}

With the representations in Tab. \ref{tab:particles-step1}, the most
general superpotential compatible with the gauge symmetry and parity
is
\begin{eqnarray} \label{eq:Wsuppot1}
W_{LR} &=& Y_Q \widehat Q \widehat \Phi \widehat Q^c 
          +  Y_L \widehat L \widehat \Phi \widehat L^c 
          - \frac{\mu}{2} \widehat \Phi \widehat \Phi
          +  f \widehat L \widehat \Delta \widehat L
          +  f^* \widehat L^c \widehat \Delta^c \widehat L^c \nonumber \\
         &+& a \widehat \Delta \widehat \Omega \widehat{\bar \Delta}
          +  a^* \widehat \Delta^c \widehat \Omega^c \widehat {\bar \Delta}^c
          + \alpha \widehat \Omega \widehat \Phi \widehat \Phi
          +  \alpha^* \widehat \Omega^c \widehat \Phi \widehat \Phi \nonumber \\
         &+& M_\Delta \widehat \Delta \widehat{\bar \Delta}
          +  M_\Delta^* \widehat \Delta^c \widehat{\bar \Delta}^c
          +  M_\Omega \widehat \Omega \widehat \Omega
          +  M_\Omega^* \widehat \Omega^c \widehat \Omega^c \, .
\end{eqnarray}
Family and gauge indices have been omitted in Eq. \eqref{eq:Wsuppot1},
more detailed expressions can be found in \cite{Aulakh:1997ba}.  Note
that this superpotential is invariant under the parity transformations
$\widehat Q \leftrightarrow (\widehat Q^c)^*$, $\widehat L
\leftrightarrow (\widehat L^c)^*$, $\widehat \Phi \leftrightarrow
\widehat \Phi^\dagger$, $\widehat \Delta \leftrightarrow (\widehat
\Delta^c)^*$, $\widehat{\bar \Delta} \leftrightarrow (\widehat{\bar
  \Delta}^c)^*$, $\widehat \Omega \leftrightarrow (\widehat
\Omega^c)^*$. This discrete symmetry reduces the number of free
parameters of the model.

The breaking of the left-right gauge group to the MSSM gauge group takes place 
in two steps: $SU(2)_R \times U(1)_{B-L} \rightarrow U(1)_R \times 
U(1)_{B-L} \rightarrow U(1)_Y$. In the first step, the neutral 
component of the triplet $\Omega$ takes a VEV,
\begin{equation}
\langle \Omega^{c \: 0} \rangle = \frac{v_R}{\sqrt{2}} \, ,
\end{equation}
which breaks $SU(2)_R$. However, since $I_{3R} (\Omega^{c \: 0}) = 0$ 
there is a $U(1)_R$ symmetry left over. Next, the group 
$U(1)_R \times U(1)_{B-L}$ is broken by
\begin{equation}
\langle \Delta^{c \: 0} \rangle = \frac{v_{BL}}{\sqrt{2}} \thickspace, \qquad 
\langle \bar{\Delta}^{c \: 0} \rangle = \frac{\bar{v}_{BL}}{\sqrt{2}} \thickspace.
\end{equation}
The remaining symmetry is now $U(1)_Y$ with hypercharge defined as
$Y = I_{3R} + \frac{B-L}{2}$. 

Regarding neutrino masses, assuming that the left triplets ($\Delta$
and $\bar \Delta$) have vanishing VEVs, one induces neutrinos masses
from a type-I seesaw only thanks to the presence of the right-handed
neutrinos \cite{Aulakh:1997ba}.

Before discussing how to test this scenario with lepton flavor
violation, let us mention some other non-minimal SUSY seesaw
models. The phenomenological study in \cite{Chao:2007ye} is based on a
model very similar to the discussed here, without $\widehat \Omega$
superfields. See also \cite{Arbelaez:2013hr} for a comprehensive study
of supersymmetric models with extended gauge groups at intermediate
steps. Finally, the seesaw mechanism can also be embedded in SUSY
GUTs, usually leading to very predictive scenarios
\cite{Calibbi:2006nq,Calibbi:2006ne,Calibbi:2009wk,Buras:2010pm,Biggio:2010me,Calibbi:2012gr}.

\subsubsection*{Probing non-minimal seesaw mechanisms}

As already discussed, a pure seesaw model predicts $A_R \simeq 0$
simply because the right sleptons do not couple to the seesaw
mediators. However, in models with non-minimal seesaw mechanisms, new
interactions between the right sleptons and the members of the
extended particle content at high energies might exist. When this is
the case, non-zero $A_R$ coefficients can be induced.

Let us consider an example. In the LR model, the left-right symmetry
implies that, above the parity breaking scale, the flavor violating
entries generated in $m_{\tilde e}^2$ are exactly as large as the ones
in $m_{\tilde L}^2$. As a consequence of this, $A_R \ne 0$ is obtained
at low energies. In fact, one can even get a handle on the symmetry
breaking pattern at high energies. Below the $SU(2)_R$ breaking scale,
parity is broken and left and right slepton soft masses evolve
differently. The left ones keep running from the $SU(2)_R$ breaking
scale to the $U(1)_{B-L}$ scale due to the left slepton couplings with
the right-handed neutrinos. One thus expects larger flavor violating
effects in the left slepton sector, and the difference between left
and right must correlate with the ratio $v_{BL}/v_R$, which measures
the hierarchy between the two breaking scales.

The question is how to measure this difference. For this purpose one
can use the positron polarization asymmetry, defined as
\begin{equation}
\mathcal{A}(\mu^+ \to e^+ \gamma) = 
\frac{|A_L|^2-|A_R|^2}{|A_L|^2+|A_R|^2} \, .
\end{equation}
If MEG observes $\mu^+ \to e^+ \gamma$ events, the angular
distribution of the outgoing positrons can be used to discriminate
between left- and right-handed polarized states and measure
$\mathcal{A}$ \cite{Okada:1999zk,Hisano:2009ae}. And this can in turn
be used to get information on $A_L$ and $A_R$.

\begin{figure}
\centering
\includegraphics[width=0.7\textwidth]{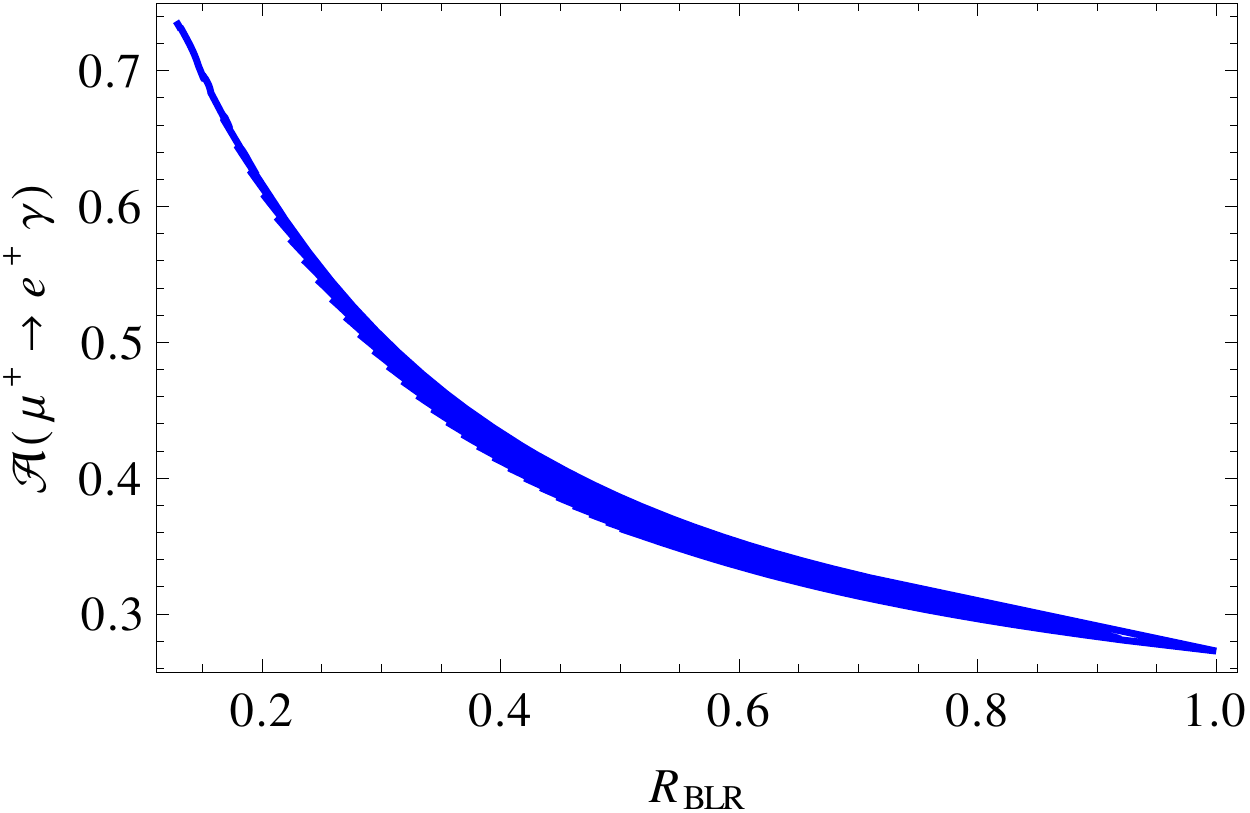}
\caption{Positron polarization asymmetry $\mathcal{A}(\mu^+ \to e^+
  \gamma)$ as a function of the ratio $R_{\text{BLR}} = \log
  (v_R/m_{\text{GUT}}) / \log (v_{BL}/m_{\text{GUT}})$. The seesaw scale $M_{SS}$
  has been fixed to $10^{13}$ GeV, whereas $v_{BL}$ and $v_R$ take
  values in the ranges $v_{BL} \in [10^{14},10^{15}]$ GeV and $v_R \in
  [10^{15},10^{16}]$ GeV. Lighter colours indicate larger
  $v_{BL}$. The CMSSM-like parameters have been taken as in the SPS3
  benchmark point \cite{Allanach:2002nj}. Figure taken from
  \cite{Esteves:2010si}.}
\label{fig:megpol}
\end{figure}

In a pure SUSY seesaw model one expects $\mathcal{A} \simeq +1$ to a
very good accuracy. However, in models with non-minimal seesaw
mechanisms $\mathcal{A}$ can significantly depart from $+1$. For
example, the LR model typically leads to significant departures from
this expectation, giving an interesting signature of the high-energy
restoration of parity. This is shown in Fig. \ref{fig:megpol},
extracted from \cite{Esteves:2010si}. First of all, it is clear than
the polarization asymmetry $\mathcal{A}(\mu^+ \to e^+ \gamma)$ is well
correlated with the quantity $\log (v_R/m_{\text{GUT}}) / \log
(v_{BL}/m_{\text{GUT}})$. One finds that as $v_{BL}$ and $v_R$ become very
different, $\mathcal{A}$ approaches $+1$. In contrast, when the two
breaking scales are close, $v_{BL}/v_R \sim 1$, this effect disappears
and the positron polarization asymmetry approaches
$\mathcal{A}=0$. Note that a negative value for $\mathcal{A}$ is not
possible in this model, since the LFV terms in the right slepton
sector never run more than the corresponding terms in the left one.

There are alternative ways to test non-minimal high-scale SUSY seesaw
scenarios. These include the study of the SUSY spectrum and, in
particular, of the \textit{invariants} pointed out for minimal seesaw
models. In this case, they contain information about the high-energy
intermediate scales \cite{DeRomeri:2011ie,Arbelaez:2013hr}

\section{Low-scale seesaw models} \label{sec:low}

The high-scale seesaw has an important drawback: the heaviness of the
seesaw mediators precludes any chance of direct tests. Only indirect
tests, based on low-energy processes which may have an imprint of the
high seesaw scale $M_{SS}$, are possible, as explained in
Sec. \ref{sec:high}. In contrast to high-scale models,
\textit{low-scale seesaw models} \cite{Boucenna:2014zba} offer a
richer phenomenological perspective since the seesaw mediators are
allowed to be light. In this type of neutrino mass models, instead of
Eq. \eqref{eq:seesaw}, neutrino masses are given by
\begin{equation} \label{eq:lowseesaw}
m_\nu \sim \mu_\nu \, \frac{v^2}{M_{SS}^2} \, ,
\end{equation}
where $M_{SS}$ is again given by the mass scale of the seesaw
mediators and $\mu_\nu$ (not to be confused with the $\mu$ parameter
of the MSSM) is a small dimensionful parameter, $\mu_\nu \ll v,
M_{SS}$. In this case, the smallness of neutrino masses is not
obtained with a large $M_{SS}$ scale, but with a tiny $\mu_\nu$
parameter. Indeed, if $M_{SS} \sim$ TeV, Eq. \eqref{eq:lowseesaw}
implies a $\mu_\nu$ parameter of the order of the eV in order to get
$m_\nu$ in the $\sim 0.1$ eV ballpark. Therefore, one can
simultaneously obtain the correct size of neutrino masses while having
seesaw mediators at the TeV scale. This leads to a plethora of new
effects, not present in high-scale seesaw models, induced by the
\textit{light} seesaw mediators. In particular, novel (and sizable)
contributions to LFV processes are possible, sometimes breaking the
relation in Eq. \eqref{eq:dipole}.

The $\mu_\nu$ parameter is intimately related to the breaking of
lepton number. In fact, in the $\mu_\nu \to 0$ limit, lepton number is
restored and the Majorana neutrino masses in Eq. \eqref{eq:lowseesaw}
vanish. This makes the smallness of the $\mu_\nu$ parameter natural,
in the sense of 't Hooft~\cite{'tHooft:1979bh}, since the symmetry of
the Lagrangian gets increased when the parameter is set to zero. For
this reason, low-scale seesaw models are also said to have
\textit{almost conserved} or \textit{slightly broken} lepton number.

The collider phenomenology of low-scale seesaw models is much richer
than that of high-scale ones. The seesaw mediators can in principle be
produced and, through their decays, one may be able to test the
mechanism behind neutrino masses. At the LHC, one typically expects
multilepton final states, often including missing energy carried away
by undetected neutrinos. In addition, the LFV signatures can be as
frequent as the flavor conserving ones. For an imcomplete list of
references on the phenomenology of low-scale seesaw models see
\cite{delAguila:2008cj,Atre:2009rg,Dev:2009aw,BhupalDev:2010he,Hirsch:2011hg,Chen:2011hc,An:2011uq,DeRomeri:2012qd,Hirsch:2012kv,Mondal:2012jv,Das:2012ze,BhupalDev:2012ru,BhupalDev:2012zg,Bandyopadhyay:2012px,Dev:2013ff,Drewes:2013gca,Helo:2013esa,Banerjee:2013fga,Dev:2013wba,Deppisch:2013cya,Aguilar-Saavedra:2013twa,Das:2014jxa,Deppisch:2015qwa,Banerjee:2015gca}. In
the case of a type-I seesaw, the seesaw mediator is a fermionic gauge
singlet. This usually suppresses its production in hadronic
colliders. However, sizable right-handed neutrino production
cross-sections are possible in some type-I seesaw realizations due to
the mixing with the left-handed neutrinos, which serves as a portal to
the gauge sector. Furthermore, when the type-I seesaw is embedded in a
left-right symmetric
scenario~\cite{Pati:1974yy,Mohapatra:1974gc,Senjanovic:1975rk} new
production mechanisms are possible thanks to the new charged currents
mediated by the $W_R^\pm$ gauge bosons. This allows for further
collider tests or the model, including searches for lepton number
violation, see for example
\cite{Keung:1983uu,AguilarSaavedra:2012fu,Das:2012ii,Chen:2013foz,Maiezza:2014ala,Bertolini:2014sua,Maiezza:2015lza}.

We now present the most popular representative of the low-scale seesaw
models: the inverse seesaw. For other low-scale seesaw models and
their LFV phenomenology see
\cite{Abada:2007ux,Abada:2008ea,Ibarra:2011xn,Dinh:2012bp,Cely:2012bz,Dinh:2013vya}.

\subsection{The supersymmetric inverse seesaw} \label{subsec:inverse}

In the supersymmetric inverse seesaw
(ISS)~\cite{Mohapatra:1986aw,Mohapatra:1986bd,Bernabeu:1987gr}, the
MSSM particle content is extended with $3$ generations of right-handed
neutrino superfields $\widehat N^c$ and $3$ generations of singlet
superfields $\widehat{X}$~\footnote{More minimal realizations of the
  ISS are possible
  \cite{Hirsch:2009ra,Malinsky:2009df,Gavela:2009cd,Dev:2012sg,Abada:2014vea}. However,
  for simplicity, we will stick to the most common version with $3+3$
  singlet superfields.}. The superpotential takes the form
\begin{equation}
 W=  W_\mathrm{MSSM} + Y_\nu \widehat N^c \widehat L \widehat H_u + M_{R} \widehat N^c \widehat X +
\frac{1}{2}\mu_{\nu} \widehat X \widehat X\,,
\label{eq:SuperPotMSSMISS}
\end{equation}
where we have omitted family indices. $Y_\nu$ and $M_R$ are general $3
\times 3$ complex mass matrices and $\mu_\nu$ is a complex symmetric
$3 \times 3$ matrix. While $M_R$ generates a lepton number conserving
Dirac mass term for the fermion singlets, $\mu_\nu$ violates lepton
number by two units. This Majorana mass term also leads to a small
mass splitting in the heavy neutrino sector, which is then composed by
three quasi-Dirac neutrinos.  The corresponding soft SUSY breaking
Lagrangian is given by
\begin{align}
-\mathcal{L}^\mathrm{soft}&=-\mathcal{L}_\mathrm{MSSM}^\mathrm{soft} 
         +   \widetilde N^c m^2_{\widetilde N}\widetilde N^{c*}
         + \widetilde X^{*} m^2_{\widetilde X} \widetilde X 
         \nonumber\\
      &
     + (T_{\nu} \widetilde N^c \widetilde L H_u +
                B_{M_R}  \widetilde N^c \widetilde X 
                +\frac{1}{2} B_{\mu_\nu}  \widetilde X \widetilde X
      + \widetilde X^{*} m^2_{\widetilde X \widetilde N} \widetilde N^c
      +\text{h.c.}) \, ,
\label{eq:softSUSY}
\end{align}
where $B_{M_R}$ and $B_{\mu_\nu}$ are the new parameters
involving the scalar superpartners of the singlet neutrino
states. Notice that while the former conserves lepton number, the
latter violates lepton number by two units.  Finally, ${\mathcal
  L}_\mathrm{MSSM}^\mathrm{soft}$ contains the soft SUSY breaking
terms of the MSSM.

The scalar potential of the model is such that the neutral components
of the Higgs superfields get non-zero VEVs,
\begin{equation}
\langle H_d^0 \rangle = \frac{v_d}{\sqrt{2}} \quad , \quad 
\langle H_u^0 \rangle = \frac{v_u}{\sqrt{2}} \, ,
\end{equation}
triggering electroweak symmetry breaking (EWSB). This induces mixings
in the neutrino sector. In the basis $\nu = (\nu_L\,,\;N^c\,,\;X)$, the
$9\times 9$ neutrino mass matrix is given by
\begin{equation}
\mathcal M_{\mathrm{ISS}}=\left(\begin{array}{c c c} 0 & m_D^T 
 & 0 \\ m_D & 0 & M_R \\ 
 0 & M_R^T & \mu_\nu \end{array}\right)\,.\label{eq:ISSmatrix}
\end{equation}
where $m_D = \frac{1}{\sqrt{2}} Y_\nu v_u$. Assuming the hierarchy
$\mu_\nu \ll m_D \ll M_R$, the mass matrix $\mathcal M_{\mathrm{ISS}}$
can be approximately block-diagonalized to give the effective mass
matrix for the light neutrinos~\cite{GonzalezGarcia:1988rw}
\begin{equation}
 m_{\mathrm{light}}\simeq m_D^T \, {M_R^T}^{-1} \mu_\nu \, M_R^{-1} \, m_D\,.
\label{LightMatrix}
\end{equation}
On the other hand, the other neutrino states form three heavy
quasi-Dirac pairs, with masses corresponding approximately to the
entries of $M_R$.

Eq. \eqref{LightMatrix} has the same form as Eq. \eqref{eq:lowseesaw},
with $M_{SS} \sim M_R$. Therefore, by taking a small $\mu_\nu$
parameter, the model allows for small neutrino masses, sizable $Y_\nu$
Yukawa couplings and singlet neutrinos at the TeV scale (or below).

\subsection{LFV in the supersymmetric inverse seesaw} \label{subsec:LFVinverse}

The presence of light singlet neutrinos induces all sorts of
effects. Here we will concentrate on their contributions to LFV
processes. For some recent works on phenomenological aspects of light
singlet neutrinos see
\cite{BhupalDev:2012zg,Abada:2012mc,Akhmedov:2013hec,Drewes:2013gca,Dev:2013wba,Dev:2013oxa,Abada:2013aba,Abada:2014nwa,Antusch:2014woa,Abada:2014cca,Deppisch:2015qwa,Antusch:2015mia,Banerjee:2015gca,Drewes:2015iva}.

There is a vast literature on LFV in models with light singlet
neutrinos. Potentially large enhancements, with respect to the usual
high-scale models, were already pointed out in early studies
\cite{Bernabeu:1987gr,Ilakovac:1994kj,Deppisch:2004fa,Deppisch:2005zm}. More
recently, several works have explored in detail the LFV anatomy of
these models, highlighting the relevance of (non-SUSY) box diagrams
induced by singlet
neutrinos~\cite{Ilakovac:2009jf,Alonso:2012ji,Dinh:2012bp,Ilakovac:2012sh},
computing Higgs penguin contributions~\cite{Abada:2011hm} and showing
enhancements in the usual photon penguin
contributions~\cite{Dev:2013oxa}. Regarding purely supersymmetric
contributions, the relevance of $Z$ penguins with right sneutrinos was
recently readdressed in \cite{Krauss:2013gya}, solving an
inconsistency in the analytical results of \cite{Arganda:2005ji} and
\cite{Hirsch:2012ax}. Finally, \cite{Abada:2014kba} constitutes the
first complete analysis of LFV in the supersymmetric inverse seesaw,
taking into account all possible contributions, supersymmetric as well
as non-supersymmetric.

We will now present the main results in \cite{Abada:2014kba}. These
were obtained using {\tt FlavorKit} \cite{Porod:2014xia}, a tool that
combines the analytical power of {\tt SARAH}
\cite{Staub:2008uz,Staub:2009bi,Staub:2010jh,Staub:2012pb,Staub:2013tta}
with the numerical routines of {\tt SPheno}
\cite{Porod:2003um,Porod:2011nf} to obtain predictions in a wide range
of models, based on the automatic computation of the lepton flavor
violating observables. See \cite{Staub:2015kfa} for a comprehensive
and pedagogical review of this set of tools.

\begin{figure}[t!]
\centering
\includegraphics[width=0.49 \linewidth]{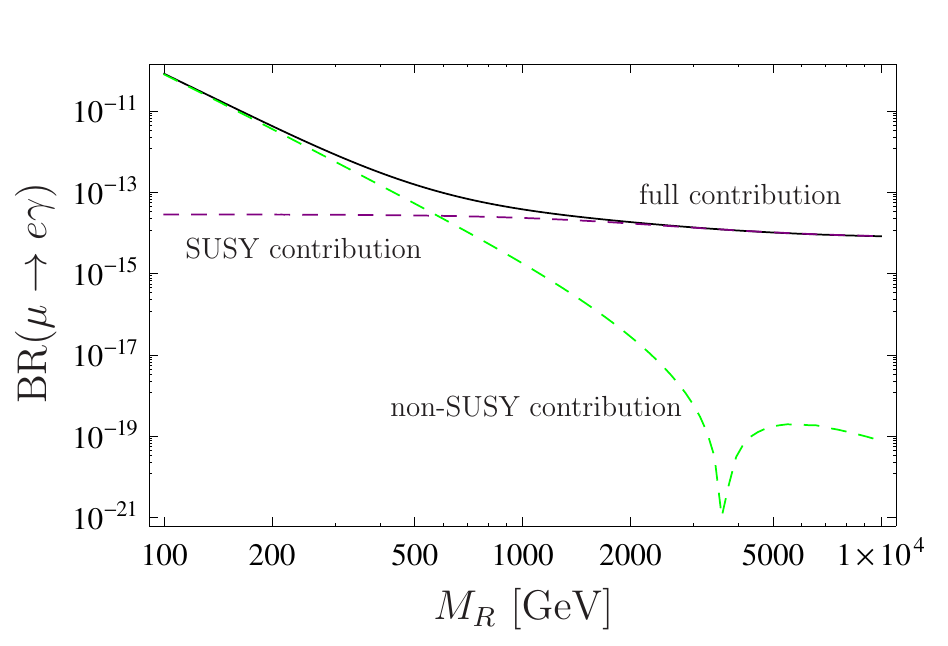}
\includegraphics[width=0.49 \linewidth]{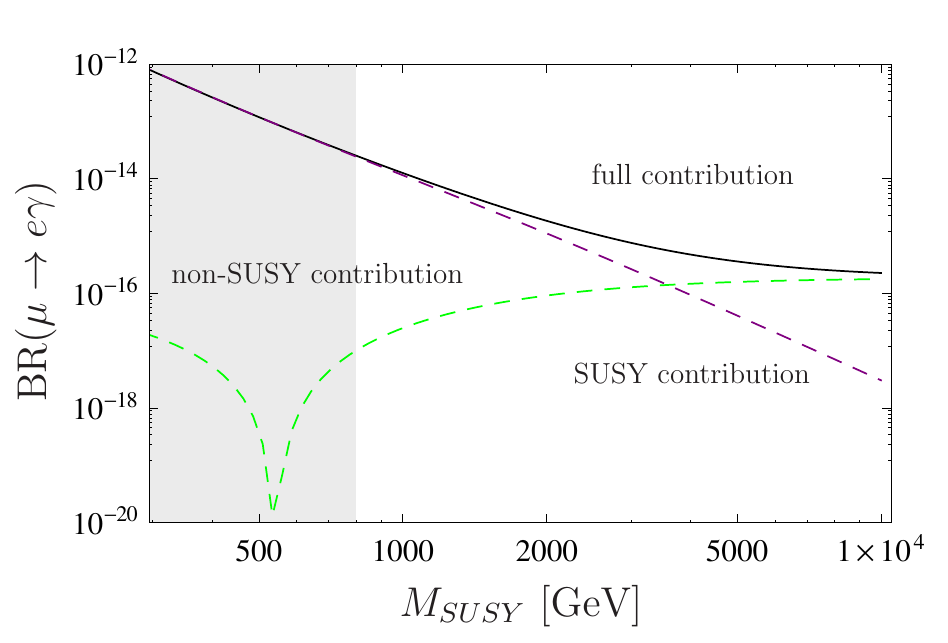}
\caption{$\text{BR}(\mu\to e \gamma)$ as a function of $M_{SUSY}$ and
  $M_R$. On the left-hand side $M_{SUSY} = 1$ TeV is fixed, whereas on
  the right-and side we set $M_R = 2$ TeV. The other parameters are
  given in the text. The gray area roughly corresponds to the
  parameter space excluded by the LHC SUSY searches. Figure taken from
  \cite{Abada:2014kba}.}
\label{fig:MuEgamma}
\end{figure}

In the following, we will discuss numerical results obtained using
universal boundary conditions at the gauge coupling unification scale,
$m_{\text{GUT}} \simeq 2 \cdot 10^{16}$ GeV, setting $M_{SUSY} = m_0 =
M_{1/2} = -A_0$. In addition, we fixed $B_{\mu_\nu} = 100 \, \mu_\nu$,
$B_{M_R} = 100 \, M_R$, $\tan \beta = 10$, $\mu > 0$ and considered a
degenerate singlet spectrum ($M_R^i \equiv M_R$ with
$i=1,2,3$). Furthermore, we fixed the $Y_\nu$ Yukawa couplings using a
modified Casas-Ibarra parameterization \cite{Casas:2001sr}, adapted
for the inverse seesaw \cite{Basso:2012ew,Abada:2012mc}, in order to
reproduce the neutrino squared mass differences and mixing angles
observed in oscillation experiments \cite{Tortola:2012te} (see also
\cite{Forero:2014bxa} for an update).

A general conclusion one can draw from \cite{Abada:2014kba} is that
the LFV phenomenology strongly depends on $M_R$ and $M_{SUSY}$. The
first scale determines the mass of the singlet neutrinos, whereas the
second one sets the superparticle masses and their relative size
determines the phenomenology. This can be seen in
Fig. \ref{fig:MuEgamma}, where BR($\mu\to e \gamma$) is shown as a
function of $M_{SUSY}$ and $M_R$. The results are displayed in three
curves: the full observable, the SUSY contributions and the non-SUSY
ones. The latter consist of contributions from $\nu$-$W^\pm$ and
$\nu$-$H^\pm$ loop diagrams, thus involving the singlet neutrinos in
combination with the $W$ boson or a charged Higgs. One finds that the
relative weight of SUSY and non-SUSY contributions is given by the
hierarchy between these two mass scales. For $M_{SUSY} \gg M_R$,
non-SUSY contributions induced by the singlet neutrinos dominate the
$\mu \to e \gamma$ amplitude, whereas for $M_{SUSY} \ll M_R$, the
usual MSSM contributions generated by chargino/sneutrino and
neutralino/slepton loops turn out to be dominant. Moreover, we find
that non-SUSY contributions can have strong cancellations.

\begin{figure}[t!]
\includegraphics[width=0.49 \linewidth]{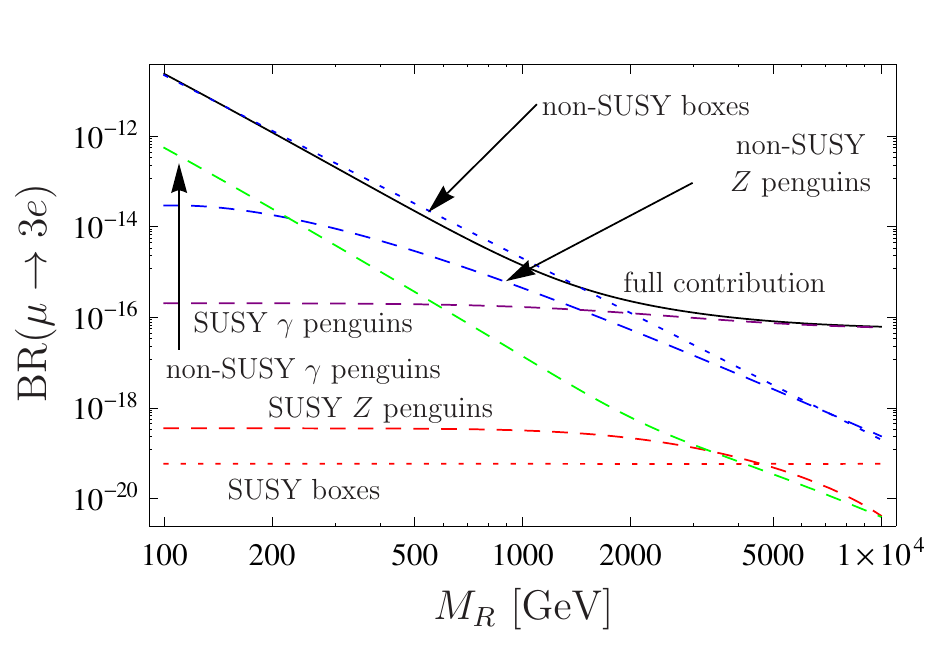}
\includegraphics[width=0.49 \linewidth]{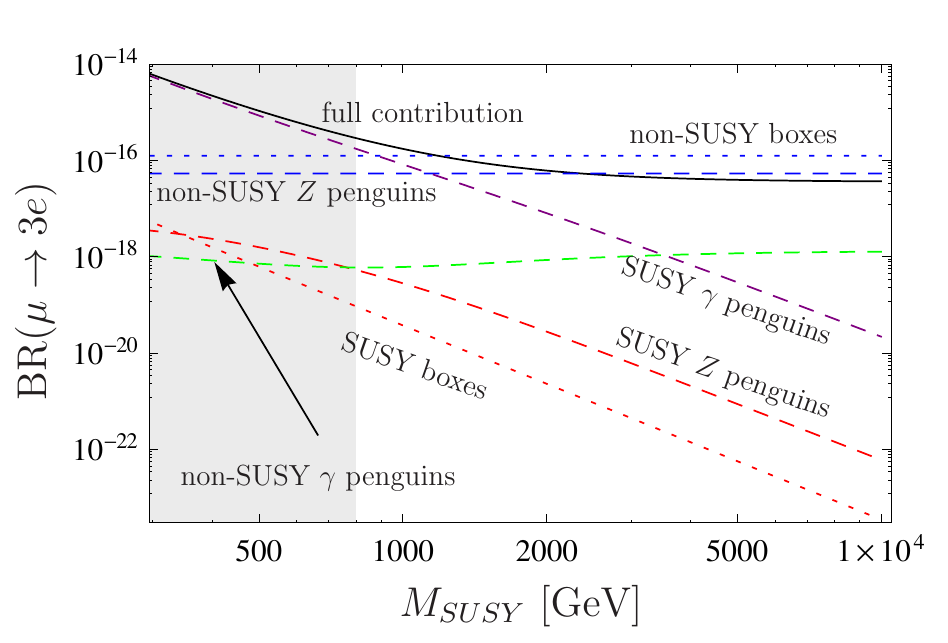}
\caption{$\text{BR}(\mu\to 3 \, e)$ as a function of $M_{SUSY}$ and
  $M_R$.  On the left-hand side $M_{SUSY} = 1$ TeV is fixed, whereas
  on the right-and side we set $M_R = 2$ TeV. The other parameters are
  given in the text. The gray area roughly corresponds to the
  parameter space excluded by the LHC SUSY searches. Figure taken from
  \cite{Abada:2014kba}.}
\label{fig:Mu3E}
\end{figure}

A similar behavior is found in case of the 3-body decays $\ell_i \to 3
\, \ell_j$. In Fig. \ref{fig:Mu3E} we display numerical results for
$\text{BR}(\mu\to 3 \, e)$ as well as for various contributions to
this observable. The anatomy of this decay is more involved, since
more types of Feynman diagrams contribute to the amplitude: SUSY and
non-SUSY photon, $Z$ and Higgs penguins, as well as box diagrams. As
for $\ell_i \to \ell_j \gamma$, we observe that non-SUSY contributions
dominate for low $M_R$ ($ < M_{SUSY}$). In particular, we see on the
left-hand side of Fig. \ref{fig:Mu3E} that non-SUSY boxes become
completely dominant as soon as one goes to $M_R$ values below $\sim 2$
TeV. This generic feature was already noted in
\cite{Ilakovac:2009jf,Alonso:2012ji,Dinh:2012bp,Ilakovac:2012sh}. For
higher values of $M_R$ SUSY contributions, and in particular the
standard photon dipole penguin, dominate. On the right-hand side we
find complementary information, with the different contributions as a
function of $M_{SUSY}$ for a fixed $M_R$. It is worth noticing
that supersymmetric $Z$ penguins never dominate.

\begin{figure}[t]
\centering
\includegraphics[width=0.7 \linewidth]{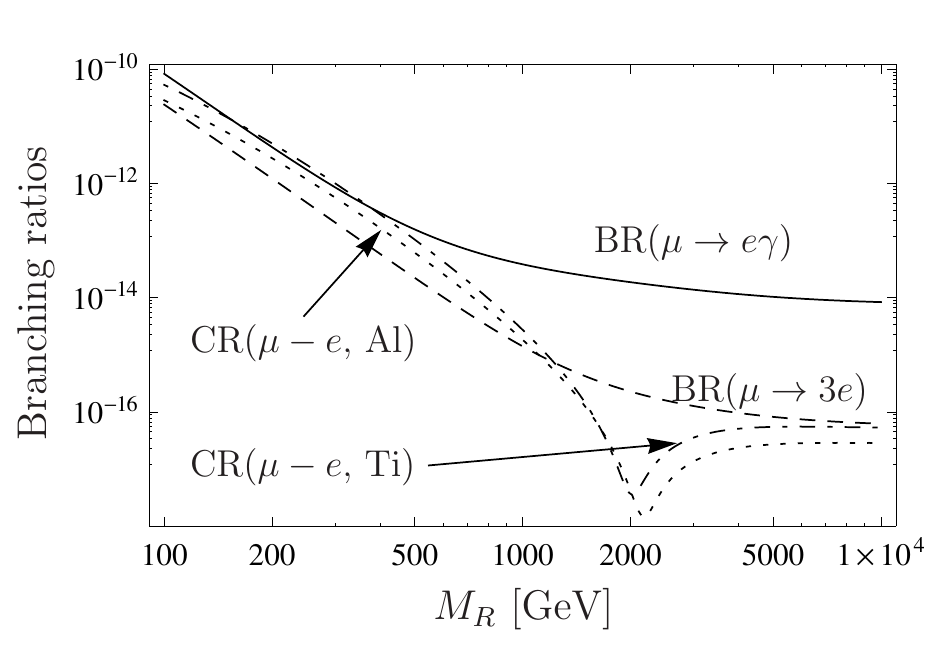}
\caption{$\text{BR}(\mu\to e \gamma)$, $\text{BR}(\mu\to 3 \, e)$ and
  $\mu-e$ conversion rates in $\mathrm{Ti}$ and $\mathrm{Al}$ as a
  function of $M_R$. $M_{SUSY}$ is fixed to $1$ TeV. The gray area
  roughly corresponds to the parameter space excluded by the LHC SUSY
  searches. Figure taken from \cite{Abada:2014kba}.}
\label{fig:MRvarious}
\end{figure}

Finally, analogous results are obtained for the $\mu-e$ conversion in
nuclei rates. Interestingly, the large non-SUSY boxes found at low
$M_R$ break the dipole dominance, leading to a clear departure from
Eq. \eqref{eq:dipole}. This is illustrated in
Fig. \ref{fig:MRvarious}, where $\text{BR}(\mu\to e \gamma)$,
$\text{BR}(\mu\to 3 \, e)$ and the $\mu-e$ conversion rates in
$\mathrm{Ti}$ and $\mathrm{Al}$ are shown as a function of
$M_R$. Indeed, for $M_R \lesssim 500$ GeV, the rates for all LFV
processes have similar sizes.  In this scenario, experiments looking
for $\mu \to 3 \, e$ and $\mu-e$ conversion in nuclei will soon
provide the most stringent constraints in this model.

\section{R-parity violating models} \label{sec:rpv}

The particle content and symmetries of the MSSM allow for the
following superpotential terms
\begin{equation}\label{rpv-superpotential}
\begin{split}
W^{\textnormal{\rpv}} = & \frac{1}{2} \lambda_{ijk} \widehat{L}_i \widehat{L}_j \widehat{e}^c_k  + \lambda'_{ijk} \widehat{L}_i \widehat{Q}_j \widehat{d}^c_k + \epsilon_i \widehat{L}_i \widehat{H}_u\\
& \: + \frac{1}{2} \lambda''_{ijk} \widehat{u}^c_i \widehat{d}^c_j \widehat{d}^c_k \, ,
\end{split}
\end{equation}
where we have explicitly introduced family indices. Here $\widehat
u^c$ and $\widehat e^c$ are the right-handed up-quark and charged
lepton superfields, respectively, and the rest of the superfields have
been already defined.  The first three terms in
$W^{\textnormal{\rpv}}$ break lepton number (L) whereas the last one
breaks baryon number (B). In principle, these couplings are not
welcome, since they give rise to many lepton and baryon number
violating processes, never observed in nature. For example, the
simultaneous presence of the $\lambda'$ and $\lambda''$ couplings
would lead to proton decay \cite{Hinchliffe:1992ad,Nath:2006ut}. This
phenomenological problem is solved in the MSSM by introducing
\textit{by hand} a new discrete symmetry that forbids all terms in
$W^{\textnormal{\rpv}}$. This symmetry is known as R-parity
\cite{Fayet:1974pd,Farrar:1978xj} and is defined as
\begin{equation}\label{Rp}
R_p = (-1)^{3(B-L)+2s} \, .
\end{equation}
Here $s$ is the spin of the particle. It is straightforward to verify
that all terms in \eqref{rpv-superpotential} break R-parity and thus
they are forbidden once this symmetry is imposed. The MSSM is defined
as R-parity conserving.

However, several arguments can be raised against R-parity:

\begin{itemize}

\item {\bf R-parity is imposed by hand.} Unlike the SM, where L and B
  conservation is automatic, in the MSSM this has to be forced by
  introducing a new symmetry, not derived from first principles. This
  is clearly a step back from the SM.
\item {\bf R-parity does not solve fast proton decay.} It is well
  known that R-parity does not forbid some dangerous dimension-5
  operators that lead to proton decay
  \cite{Ibanez:1991pr,Sakai:1981pk,Weinberg:1981wj}. For example, the
  operator $\mathcal{O}_5 = \frac{f}{M} \widehat Q \widehat Q \widehat
  Q \widehat L$ has $R_p(\mathcal{O}_5)=+1$ and thus conserves
  R-parity. The bounds obtained from the non-observation of proton
  decay imply that, even for $M = M_{\text{Planck}}$, $f$ must be
  smaller than $10^{-7}$~\cite{Ellis:1983qm}. In order to forbid
  $\mathcal{O}_5$ and other similar dimension-5 operators, one may
  resort to additional flavor symmetries \cite{Morisi:2012hu}.
\item {\bf There is no reason to forbid all the L and B violating
  operators.} Proton decay requires the simultaneous pesence of L and
  B violating couplings. Therefore, it is sufficient to impose the
  conservation of just one of these two symmetries in order to forbid
  proton decay. This has led to the consideration of alternative
  discrete symmetries which allow for either L or B violation while
  protecting the proton. An example of such symmetries is baryon
  triality ($Z_3^B$) \cite{Ibanez:1991pr,Dreiner:2005rd}.

\end{itemize}

Furthermore, there are several good motivations to consider R-parity
violating (\rpvn) scenarios. The violation of lepton number by any of
the first three couplings in Eq. \eqref{rpv-superpotential}
automatically leads to non-zero neutrino masses
\cite{Hall:1983id,Ross:1984yg,Hirsch:2004he}. Moreover, the presence
of \rpv couplings leads to a rich collider phenomenology due to the
decay of the LSP. This can be translated into longer decay chains,
changing the expected signatures at the LHC
\cite{Dreiner:2012wm,Dreiner:2012ec}. In fact, \rpv has also been
considered as a way to relax the stringent bounds on the squark and
gluino masses, see for example
\cite{Graham:2012th,Hanussek:2012eh,Evans:2012bf,Bhattacherjee:2013gr}.

Finally, in \rpv the standard neutralino LSP is lost as a dark matter
candidate. Therefore, alternative candidates must be
considered. Examples in the literature include (i) gravitinos
\cite{Borgani:1996ag,Takayama:2000uz,Hirsch:2005ag}, (ii) the axion
\cite{Kim:1986ax,Raffelt:1996wa} or (iii) its superpartner, the axino
\cite{Chun:1999cq,Chun:2006ss}. For general reviews on R-parity
violation and collections on bounds on the \rpv couplings see
\cite{Barbier:2004ez,Chemtob:2004xr,Kao:2009fg,Mohapatra:2015fua}.

We will now discuss separately the LFV phenomenology of two very
different supersymmetric scenarios with R-parity violation: explicit
R-parity violation (e-\rpvn) and spontaneous R-parity violation
(s-\rpvn).

\subsection{Explicit R-parity violation} \label{subsec:ERPV}

The most characteristic signatures of R-parity violating models are,
of course, processes with L or B violation. Nevertheless, processes
that violate lepton flavor can provide interesting signatures as well
and, in fact, they can be more attractive due to the large number of
upcoming LFV experiments.

\subsubsection*{Higgs LFV decays}

After the historical discovery of the Higgs
boson~\cite{Aad:2012tfa,Chatrchyan:2012ufa}, a lot of effort has been
put into the determination of its properties. In particular, the Higgs
boson decays may contain a lot of valuable information, with potential
indications of new physics. Recently, the CMS collaboration reported
on an intriguing $2.4 \sigma$ excess in the $h\to \tau \mu$
channel~\cite{Khachatryan:2015kon}. This hint, which translates into
$\text{BR}(h\to\tau\mu) = \left( 0.84_{-0.37}^{+0.39} \right)$, is
based on the analysis of the 2012 dataset, taken at $\sqrt{s}=8$~TeV
and an integrated luminosity of 19.7~fb$^{-1}$. This large Higgs
LFV~\footnote{For pioneer works on Higgs LFV decays see
  \cite{Pilaftsis:1992st,DiazCruz:1999xe}.}  branching ratio is quite
challenging and most NP models cannot accommodate it
\cite{Dorsner:2015mja}. In fact, the flavor conserving Higgs decay $h
\to \tau \tau$ has a branching ratio of \textit{only} $\sim 6 \%$, not
much higher than the LFV one found by CMS. Although independent
confirmation by ATLAS, as well as additional statistics in CMS, would
be required in order to promote this hint to the category of evidence
of new physics, it is interesting to explore different models in order
to determine what type of frameworks can accommodate this signal.

\begin{figure}
\centering
\vspace{5mm}
\includegraphics[width=0.38\textwidth]{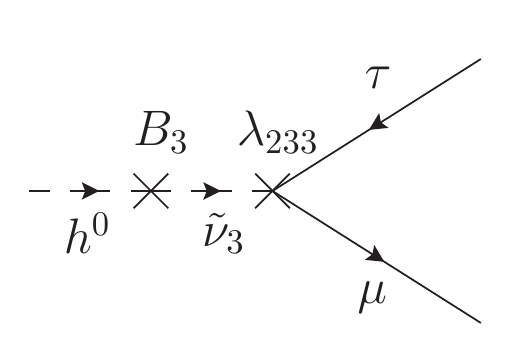}
\hspace*{5mm}
\includegraphics[width=0.38\textwidth]{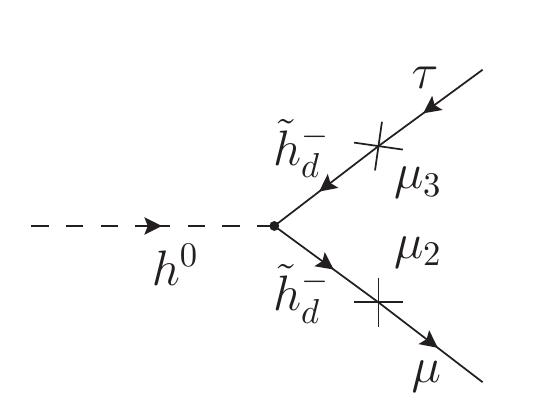}
\caption{Tree-level \rpv contributions to $h \to \tau \mu$. On the
  left, a $B \, \lambda$ contribution. On the right, a $\epsilon_i^2$
  (denoted as $\mu_i^2$ in \cite{Arhrib:2012mg}) contribution. Figure
  borrowed from \cite{Arhrib:2012mg}.}
\label{fig:contributions}
\end{figure}

\begin{figure}
\centering
\vspace{5mm}
\includegraphics[width=0.7\textwidth]{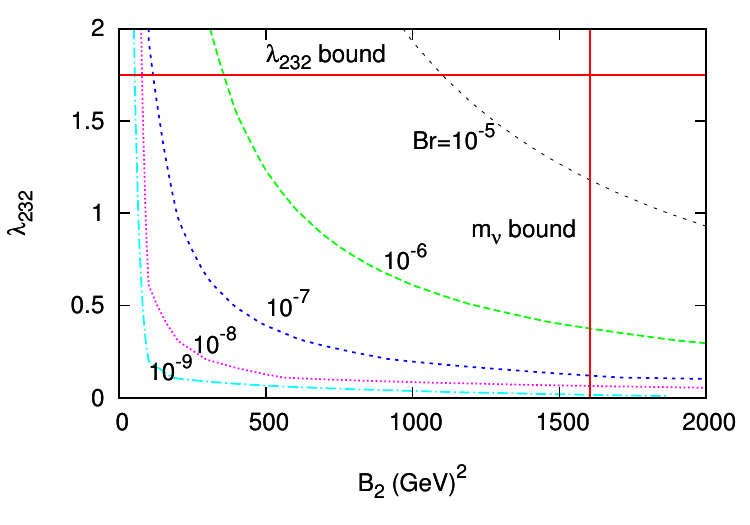}
\caption{$\text{BR}(h \to \mu \tau)$ contours on the $B_2 -
  \lambda_{232}$ plane. The continuous horizontal and vertical lines
  show approximate limits due to neutrino masses (in case of $B_2$)
  and charged current experiments (in case of $\lambda_{232}$). Figure
  borrowed from \cite{Arhrib:2012mg}.}
\label{fig:resulthmutau}
\end{figure}

Regarding supersymmetric models, several scenarios have been recently
explored, some of them even before the CMS hint was announced. In
particular, the authors of \cite{Arhrib:2012mg,Arhrib:2012ax}
considered an extension of the MSSM including all L violating
couplings in Eq. \eqref{rpv-superpotential}.  The particles-sparticles
mixing due to the \rpv couplings induce Higgs LFV decays at
tree-level, thus potentially being able to reach branching ratios as
high as the one found by the CMS collaboration.  Two specific examples
are shown in Fig. \ref{fig:contributions}. However, the existing
experimental bounds on the relevant combinations of \rpv couplings
contributing to $h \to \tau \mu$ forbid such large LFV branching
ratios. For example, in the $B \lambda$ contribution both \rpv
parameters are strongly constrained. In case of $B_i$, the \rpv
mixings between the Higgs boson and the sneutrinos ($\mathcal{L}
\supset B_i \widetilde L_i H_u$), they have strong bounds since they
induce non-zero neutrino
masses~\cite{Grossman:1998py,Chun:1999bq,Davidson:2000uc}. Moreover,
the $\lambda$ couplings are constrained by charged current experiments
\cite{Barbier:2004ez}. Once these constraints are taken into account,
the maximum $\text{BR}(h \to \tau \mu)$ one can get is not very
impressive. This is illustrated in Fig. \ref{fig:resulthmutau}, where
$\text{BR}(h \to \tau \mu)$ contours are drawn on the $B_2 -
\lambda_{232}$ plane. From this figure one concludes that $\text{BR}(h
\to \tau \mu)$ can reach, at most, $\text{a few} \, \times \,
10^{-5}$, clearly below the CMS hint. Similar conclusions are obtained
when other combinations of \rpv parameters are considered.

The supersymmetric inverse seesaw has also been considered as a
possible setup to reproduce a Higgs LFV branching ratio into $\tau
\mu$ at the $1 \%$ level \cite{Arganda:2014dta}. In this case, $h \to
\tau \mu$ takes place at 1-loop, naturally suppressing the branching
ratio. As a result of this, as well as due to the constraints from
other LFV processes such as $\ell_i \to \ell_j \gamma$, one finds that
the maximum allowed $\text{BR}(h \to \tau \mu)$ is about $\sim
10^{-5}$. Therefore, this model cannot account for a branching ratio
as obtained by CMS either. In order to conclude this discussion on $h
\to \tau \mu$ with a positive note, let us mention that known models
that can account for $\text{BR}(h \to \tau \mu) \sim 1 \%$ exist in
the literature. They all involve extended Higgs sectors. In
particular, it has been shown that Two-Higgs-Doublet models of
type-III, in which both Higgs doublets can couple to up- and down-type
fermions, can easily accommodate the CMS signal
\cite{Davidson:2010xv,Kopp:2014rva,Sierra:2014nqa,Heeck:2014qea,Crivellin:2015mga,Dorsner:2015mja}. The
MSSM, being a Two-Higgs-Doublet models of type-II, cannot~\footnote{In
  fact, several studies have shown that one cannot accommodate
  $\text{BR}(h \to \tau \mu) \sim 1 \%$ when the low-energy theory is
  the MSSM~\cite{Arganda:2004bz,Arana-Catania:2013xma}. The same
  conclusion applies to heavy Higgs LFV
  decays~\cite{DiazCruz:2008ry}.}.

\subsubsection*{Trilinear R-parity violation and LFV}

\begin{figure}
\centering
\vspace{5mm}
\includegraphics[width=0.48\textwidth]{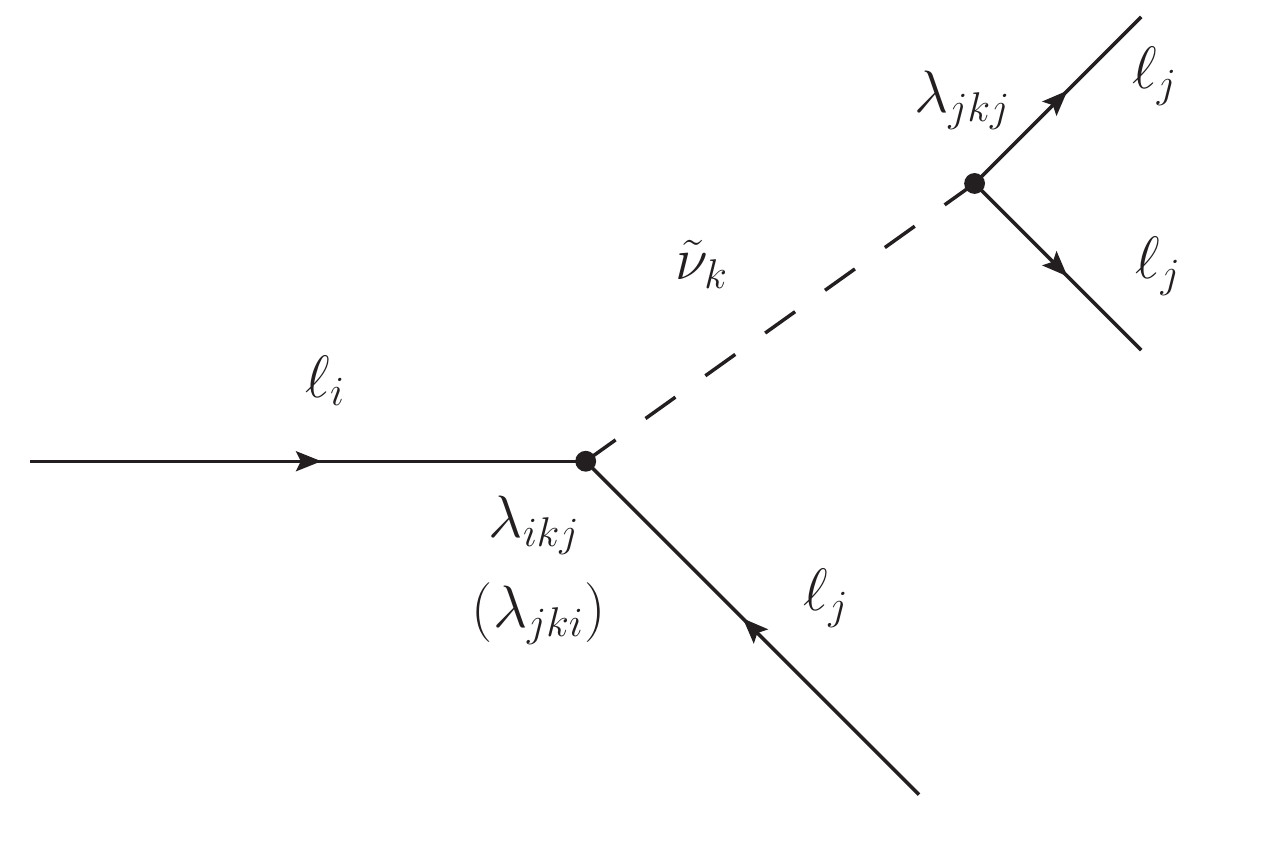}
\includegraphics[width=0.48\textwidth]{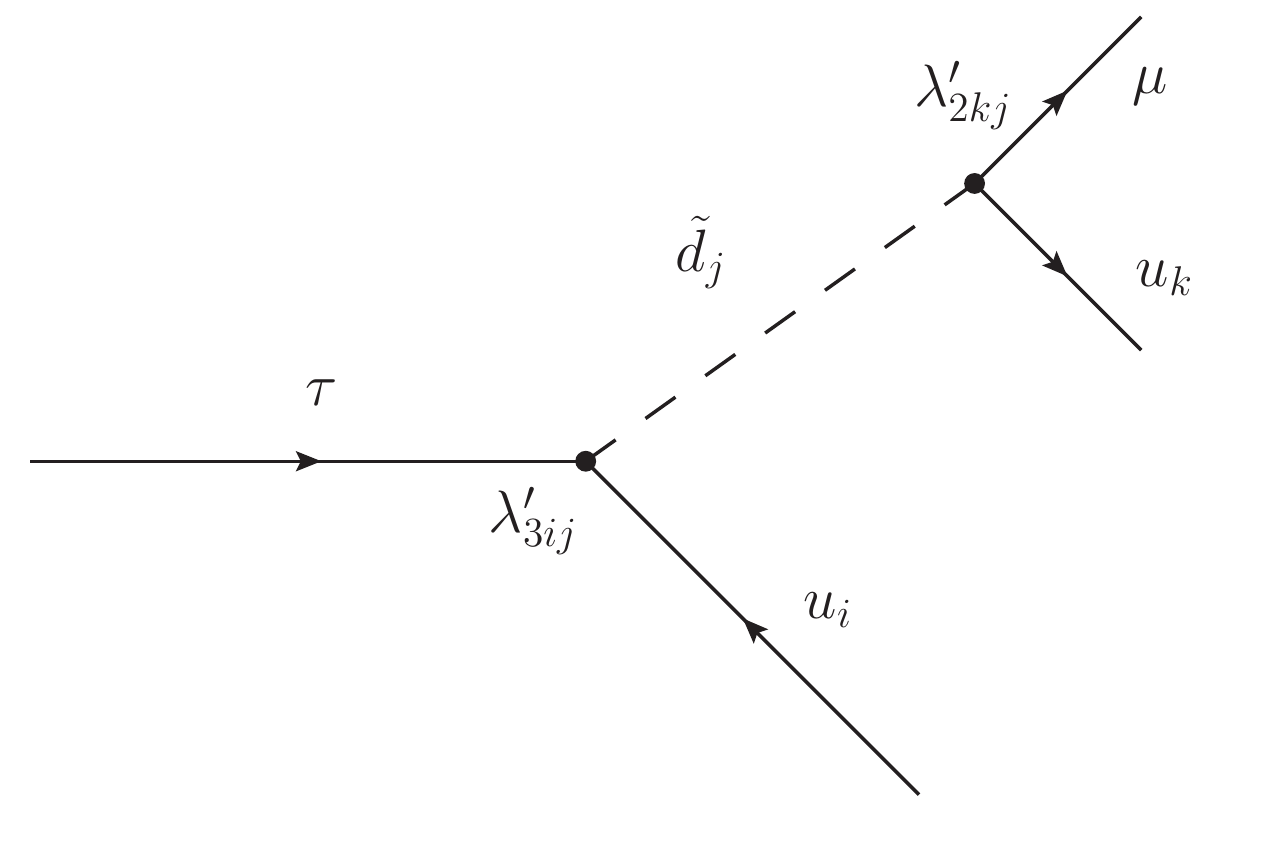}
\caption{To the left, $\ell_i \to 3 \, \ell_j$ induced by trilinear
  $\lambda$ couplings and sneutrino exchange. To the right, $\tau \to
  M \mu$ induced by $\lambda'$ couplings and squark exchange.}
\label{fig:L3Ltree}
\end{figure}

In principle, the usual LFV processes studied in R-parity conserving
models can be studied in the R-parity violating ones and, in some
cases, they get additional \rpv contributions. This is the case of
$\ell_i \to 3 \, \ell_j$, $M \to \ell_i \ell_j$ and $\tau \to M
\ell_i$, where $M$ is a neutral meson, which, in the presence of
trilinear \rpv couplings, can be induced at tree-level
\cite{deGouvea:2000cf}. This is represented in Fig. \ref{fig:L3Ltree},
where two examples are shown: $\ell_i \to 3 \, \ell_j$ induced by
$\lambda$ couplings and sneutrino exchange and $\tau \to M \mu$
induced by $\lambda'$ couplings and squark exchange. This allows one
to derive a large collection of bounds on the size of the trilinear
couplings and the masses of the superparticles mediating the LFV
decays \cite{Dreiner:2006gu,Daub:2012mu}.

Let us consider the right-hand side of Fig. \ref{fig:L3Ltree}. After
dressing the quarks in the final state, this Feynman diagram induces
$\tau \to M \mu$ at tree-level. Since the exchanged particle, a
down-type squark in this case, is much heavier than the rest of
particles, this process can be well described by the 4-fermion
effective Lagrangian
\begin{equation}
\mathcal L_{\text{eff}} = - \frac{\lambda'_{3ij} \lambda_{2kj}^{' \ast}}{m_{\tilde d_j}^2} \, \left( \overline{\tau^c} P_L u_i \right) \, \left( \bar u_k P_R \mu^c \right) \, ,
\end{equation}
obtained after integrating out the heavy squark. One can now use this
Lagrangian and, together with the relevant hadronic form factors,
compute rates for processes such as $\tau \to \mu \pi^+ \pi^-$. The
authors of \cite{Daub:2012mu} followed this method and used Belle
results on searches for $\tau$ LFV decays
\cite{Miyazaki:2008mw,Miyazaki:2011xe,Miyazaki:2013yaa} to obtain the
limit
\begin{equation}
\lambda'_{31j} \lambda_{21j}^{' \ast} < 2.1 \cdot 10^{-4} \, \left( \frac{m_{\tilde d_j}}{100 \, \text{GeV}} \right)^2 \, .
\end{equation}

One can exploit this idea using other LFV observables
involving mesons. We refer to \cite{Dreiner:2006gu,Daub:2012mu} for a
more complete list of constraints.

Similarly, trilinear \rpv couplings can also trigger $\mu-e$
conversion in nuclei, induced by diagrams very similar to the one on
the right-hand side of Fig. \ref{fig:L3Ltree}. Interestingly, in this
case $\mu-e$ conversion in nuclei would take place at tree-level,
while the more popular $\mu \to e \gamma$ would take place at
1-loop. This has been recently pointed out by the authors of
\cite{Sato:2014ita}, who argue that experiments looking for $\mu-e$
conversion in nuclei might be the first (and perhaps the only ones) to
observe a non-zero signal in the next round of experiments.

\subsubsection*{Other results on LFV in \rpv scenarios}

\begin{figure}
\centering
\vspace{5mm}
\includegraphics[width=0.25\textwidth]{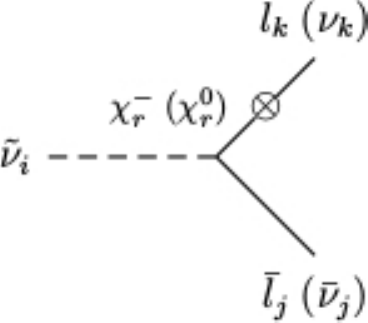}
\caption{Sneutrino decay to $\ell_k \bar \ell_j$ in bilinear R-parity
  violation. The open circle with a cross indicates the \rpv induced
  mixing between charginos and charged leptons. Figure taken from
  \cite{AristizabalSierra:2012qa}.}
\label{fig:SnuDecay}
\end{figure}

Before concluding, let us briefly comment on other aspects of LFV in
\rpv models. An interesting feature of \rpv models is that some lepton
number violating processes at colliders might look like lepton flavor
violating ones. This is for example the case of sneutrino decay in
bilinear R-parity violation \cite{AristizabalSierra:2012qa}, as shown
in Fig. \ref{fig:SnuDecay}. This process is possible thanks to the
mixing between the MSSM charginos and the standard charged leptons. At
the LHC, if the sneutrinos are directly produced, the absence of
missing energy would make this process look like a LFV one.

Finally, for other recent works on LFV in \rpv models see
\cite{Zhang:2013hva,Zhang:2013jva,Wang:2014dba}.

\subsection{Spontaneous R-parity violation} \label{subsec:SRPV}

A very attractive scenario for LFV is that of spontaneous R-parity
violation. When a scalar field oddly charged under R-parity gets a VEV
in a theory with a R-parity conserving Lagrangian, R-parity gets
spontaneous broken. Here we will concentrate on spontaneous L and
$R_p$ violation. Although R-parity is a discrete symmetry, its
breaking comes along with the breaking of the continuous global
symmetry $U(1)_L$. This implies the existence of a massless Goldstone
boson, usually called the majoron ($J$)
\cite{Chikashige:1980ui,Gelmini:1980re}.

The nature of the majoron is crucial for the phenomenological success
of the model. In fact, in the first model with s-\rpv
\cite{Aulakh:1982yn}, the breaking of R-parity was triggered by the
VEV of a left-handed sneutrino. This simple setup was eventually
excluded since the doublet nature of the majoron leads to conflict
with LEP bounds on the $Z$ boson invisible decay width and
astrophysical data \cite{Raffelt:1996wa,Amsler:2008zzb}. However,
more refined models where the violation of lepton number is induced by
a gauge singlet are perfectly valid possibilities. As a benchmark
example of this family we will consider here the model introduced in
\cite{Masiero:1990uj}. For alternative models with gauged lepton
number see for example
\cite{Hayashi:1984rd,FileviezPerez:2008sx,Barger:2008wn}.

In the model of Ref. \cite{Masiero:1990uj}, the particle content is
extended with three additional singlet superfields, namely,
$\widehat\nu^c$, $\widehat S$ and $\widehat\Phi$, with lepton number
assignments of $L=-1,1,0$ respectively. By assumption, the Lagrangian
of the theory conserves lepton number. Therefore, the superpotential
can be written as
\begin{equation}
W_{SRPV} = W_\mathrm{MSSM} + Y_\nu \widehat L \widehat N^c \widehat H_u
          - h_0 \widehat H_d \widehat H_u \widehat\Phi
          + h \widehat\Phi \widehat N^c \widehat S +
          \frac{\lambda}{3!} \widehat\Phi^3 \, .
\label{eq:Wsuppot}
\end{equation}
For simplicity, one can simply consider one generation of
$\widehat N^c$ and $\widehat S$ superfields. Several scalar fields
acquire VEVs after electroweak symmetry breaking. In addition to the
usual MSSM Higgs boson VEVs, $v_d$ and $v_u$, these are $\langle \Phi
\rangle = v_{\Phi}/\sqrt{2}$, $\langle {\widetilde N}^c \rangle =
v_R/\sqrt{2}$, $\langle {\widetilde S} \rangle = v_S/\sqrt{2}$ and
$\langle \widetilde{\nu_L}_i \rangle = v_i/\sqrt{2}$.  This vacuum
configuration breaks lepton number and R-parity. In fact, we notice
that $v_R \ne 0$ generates the effective bilinear \rpv terms
$\epsilon_i = Y_\nu^i v_R/\sqrt{2}$. Furthermore, neglecting $v_i \ll
v_R,v_S$, one finds the resulting majoron profile
\begin{equation}\label{eq:majoron}
J \simeq \text{Im} \left( \frac{v_S}{V} \, \widetilde S -\frac{v_R}{V} \, \widetilde N^c \right) \, ,
\end{equation}
where $V=\sqrt{v_R^2+v_S^2}$. Eq. \eqref{eq:majoron} shows that the
majoron inherits the singlet nature of the scalar fields that break
lepton number with their VEVs, thus suppressing the couplings to the
$Z$ boson and evading the stringent LEP bound.

Here we are interested in novel LFV features due to the presence of
the majoron~\footnote{Another interesting signature present in majoron
  models is the invisible decay of the Higgs boson, $h \to J J$
  \cite{Joshipura:1992hp,Hirsch:2004rw}.}. This new massless state
dramatically changes the phenomenology both at collider and low-energy
experiments \cite{Hirsch:2008ur,Hirsch:2009ee}. In particular, it
leads to new LFV processes, such as $\mu \to e J$ or $\mu \to e J
\gamma$. The exotic muon decay $\mu \to e J$ was first studied in
\cite{Romao:1991tp} and later revisited in \cite{Hirsch:2009ee}, where
the decay with an additional photon was also considered.  Furthermore,
the impact of the majoron on $\mu-e$ conversion in nuclei was
discussed in \cite{GarciaiTormo:2011et}~\footnote{See also
  \cite{Celis:2014iua} for similar LFV processes in the context of
  invisible axions.}.

The rate of the $\mu \to e J$ decay is determined by the $e-\mu-J$
coupling, $O_{e \mu J}$, which, in the model under consideration, is
of the form $O_{e \mu J} \sim \frac{1}{v_R} \, \times \, \text{RPV
  parameters}$. This makes us conclude that, in general, one expects
large partial muon decay widths to majorons for low $v_R$. However,
currently there are no experiments looking for $\mu \to e J$ and the
current best limit on the branching ratio, $\text{BR}(\mu \to e J)
\lesssim 10^{-5}$, dates back to 1986 \cite{Jodidio:1986mz}. Regarding
the decay including a photon, $\mu \to e J \gamma$, one can profit
from the MEG experiment and its search for the more popular channel
$\mu \to e \gamma$.

The two branching ratios are related by
\begin{equation}
\text{BR}(\mu \to e J \gamma) = \frac{\alpha}{2 \pi} {\cal I}(x_{min},y_{min}) \, \text{BR}(\mu \to e J) \, .
\end{equation}
Here ${\cal I}(x_{min},y_{min})$ is a 3-body phase space integral
defined as
\begin{equation} \label{eq:phase}
{\cal I}(x_{min},y_{min}) = \int dx dy \frac{(x-1)(2-xy-y)}{y^2(1-x-y)} \, ,
\end{equation}
the dimensionless parameters $x$, $y$ are defined as
\begin{equation}
x = \frac{2 E_e}{m_\mu} \quad , \quad y = \frac{2 E_\gamma}{m_\mu}
\end{equation}
and $x_{min}$ and $y_{min}$ are the minimal electron and photon
energies that a given experiment can measure. Indeed, the integral in
Eq. \eqref{eq:phase}, which would contain infrarred and collinear
divergences, is regularized by the specific choices made by an
experiment.

\begin{figure}
\centering
\vspace{5mm}
\includegraphics[width=0.48\textwidth]{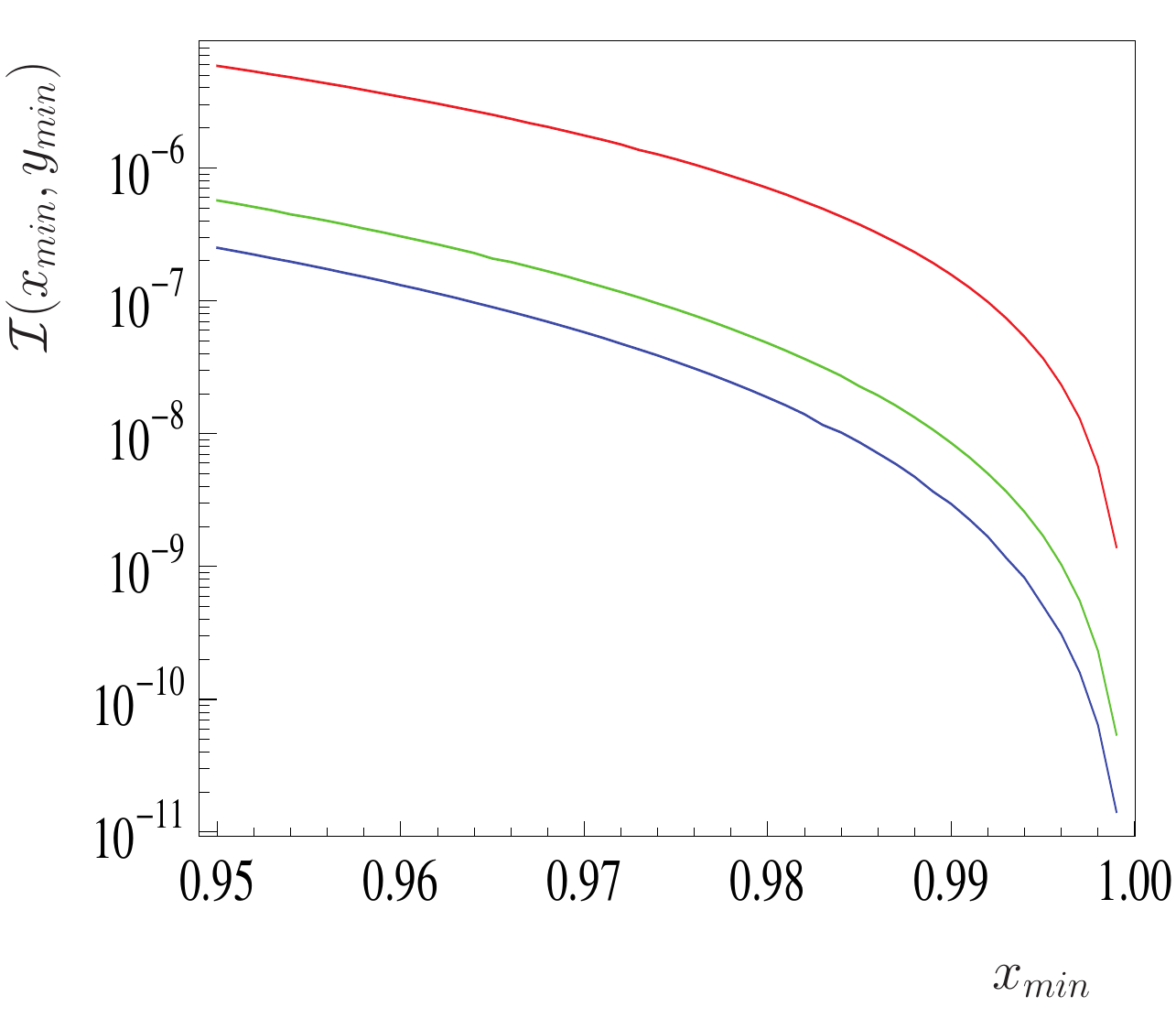}
\includegraphics[width=0.48\textwidth]{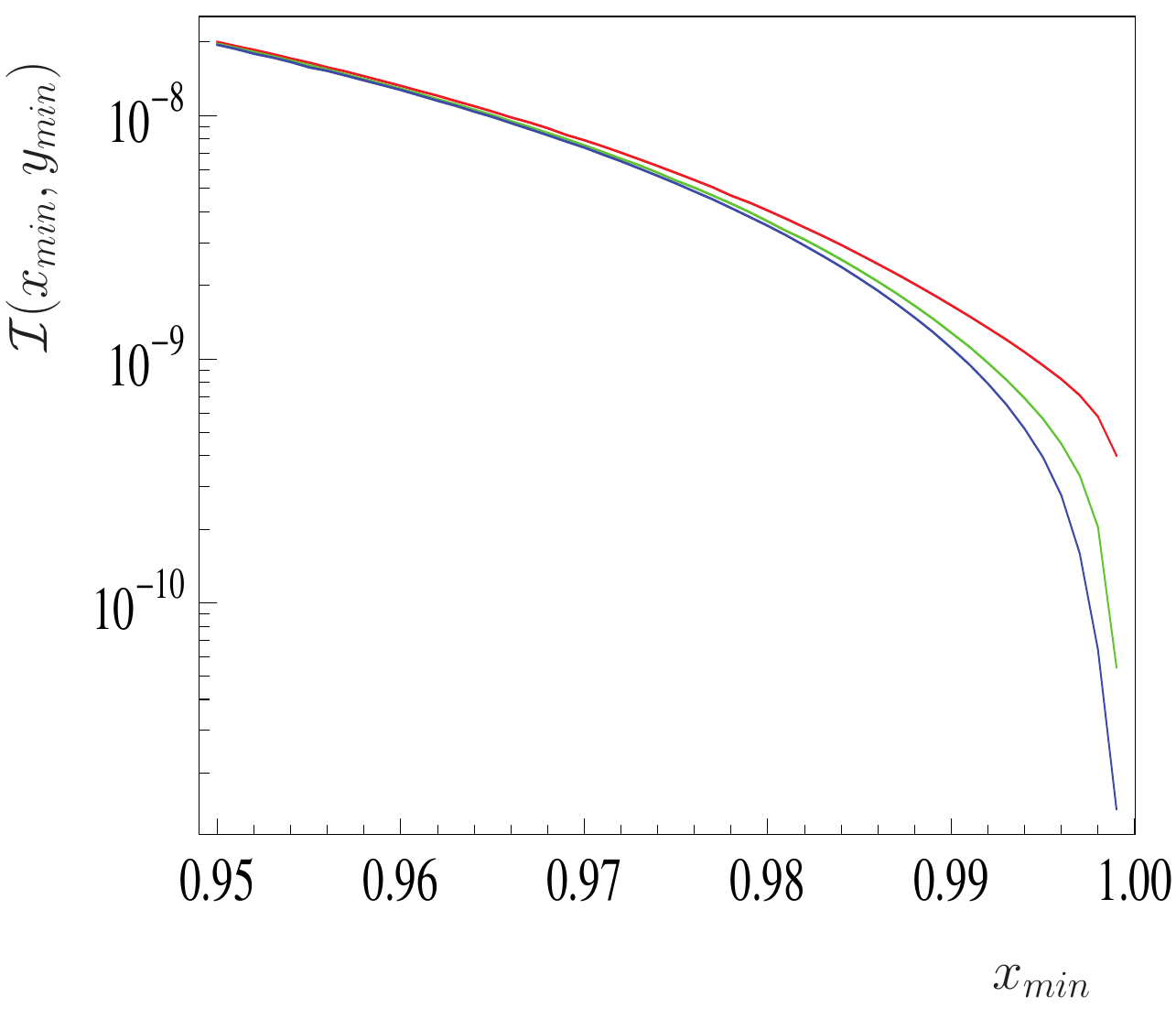}
\caption{The phase space integral for the decay $\mu\to e J \gamma$ as
  a function of $x_{min}$ for three different values of $y_{min}=
  0.95, 0.99, 0.995$ from top to bottom and for two different values
  of $\cos\theta_{e\gamma}$. To the left $\cos\theta_{e\gamma}=-0.99$,
  whereas to the right $\cos\theta_{e\gamma}=-0.99997$. Figure taken
  from \cite{Hirsch:2009ee}.}
\label{fig:int}
\end{figure}

As explained above, the main advantage of $\mu \to e J \gamma$ is the
existence of the MEG experiment. However, the question is whether it
is sensitive to this exotic LFV process or not. Fig. \ref{fig:int}
shows the value of the phase space integral ${\cal
  I}(x_{min},y_{min})$ as a function of $x_{min}$ for three different
values of $y_{min}$ and for two choices of $\cos\theta_{e\gamma}$ (the
relative angle between the electron and photon
directions). Unfortunately, the MEG experiment is specifically
designed for a single search. In fact, the cuts used in the search for
$\mu\to e \gamma$ are very restrictive: $x_{min}\ge 0.995$,
$y_{min}\ge 0.99 $ and $|\pi - \theta_{e\gamma}| \le$ 8.4 mrad. For
these exact values one finds a tiny phase space integral, ${\cal I}
\simeq 6 \cdot 10^{-10}$. As a consequence of this, a limit for
$\text{BR}(\mu\to e \gamma)$ of the order of $\le 10^{-13}$ would
translate into the useless limit $\text{BR}(\mu\to e J) < 0.14$. To
improve upon this bound, it is necessary to relax the cuts. For
example, by relaxing the cut on the opening angle to
$\cos\theta_{e\gamma}=-0.99$. However, this is prone to induce
additional unwanted background events. In particular, accidental
background from muon annihilation in flight. Therefore, although one
could in principle increase the value of the phase space integral
${\cal I}(x_{min},y_{min})$, the background in that case would make
the search for a positive signal impossible. This discussion suggests
that a better timing resolution of the experiment would be welcome in
order to reduce the background and be sensitive to final states
including majorons.

\section{Summary and conclusions} \label{sec:conclusions}

In summary, we have reviewed the lepton flavor violating phenomenology
of several non-minimal supersymmetric models: high-scale and low-scale
seesaw models as well as models with explicit or spontaneous R-parity
violation. The main conclusion from this overview is that the lepton
flavor violating signatures can be very different from those found in
the MSSM. This translates into two important messages:

\begin{itemize}
\item {\bf For the theorists:} Lepton flavor violation might be much
  more intricate than what minimal models predict. Therefore, we
  should be careful when extrapolating our expectations (derived from
  the MSSM) to extended frameworks.
\item {\bf For the experimentalists:} Although minimal models are of
  course well motivated, lepton flavor violation might show up in
  non-standard channels. We must be ready to avoid missing a relevant
  signal.
\end{itemize}

Properly identifying the underlying physics will be crucial
in case a positive observation in one or several LFV experiments is
made. This problem might be soon have to be addressed, given the
exciting projects that are currently going on or soon starting their
search for LFV. Hopefully, this review, as well as the many
phenomenological studies in the bibliography, will help shedding some
light on this matter.

\section*{Acknowledgements}

I am extremely grateful to my collaborators in the subjects discussed
in this review. I also acknowledge partial support from the
EXPL/FIS-NUC/0460/2013 project financed by the Portuguese FCT.

\bibliographystyle{utphys}
\bibliography{LFV}

\end{document}